\documentclass[12pt]{article}
\usepackage[english]{babel}
\usepackage[utf8]{inputenc}
\usepackage[T1]{fontenc}
\usepackage{bold-extra}
\usepackage{subfig}
\usepackage[subfigure]{tocloft}
\usepackage{comment}
\usepackage{amsfonts, amsmath, amsthm, amssymb}
\usepackage{amscd}
\usepackage{amsfonts}
\usepackage{anonchap}
\usepackage{mathtools}
\usepackage[hang, flushmargin]{footmisc}
\usepackage{footnotebackref}
\usepackage[table,xcdraw]{xcolor}
\usepackage{extsizes}
\usepackage[final]{graphicx}
\usepackage{float}
\usepackage{physics}
\usepackage{pdfpages}
\usepackage{slashed}
\usepackage[title,toc,titletoc]{appendix}
\usepackage{titlesec}
\usepackage{tocloft}
\usepackage{bbm}
\usepackage{tikz-cd}
\definecolor{rossoCP3}{cmyk}{0,.88,.77,.40}
\usetikzlibrary{decorations.pathmorphing}
\usepackage[sorting=none, style=nature]{biblatex}
\usepackage{ytableau}
\usepackage{youngtab}
\usepackage{listings}
\usepackage[margin=0.8in]{geometry}
\usepackage{xcolor}
\usepackage{footmisc}
\usepackage{footnotebackref}
\usepackage{hyperref}
\hypersetup{
    colorlinks,
    citecolor=black,
    filecolor=black,
    linkcolor=black,
    urlcolor=black
}

\titleformat{\section}
{\large\bfseries\scshape\centering}{\thesection\ }{1em}{}

\titleformat{\subsection}
  {\normalsize\bfseries\boldmath\centering}{\thesubsection}{1em}{}

\renewcommand\thesection{\arabic{section}.\ }  
  \renewcommand\thesubsection{\arabic{section}.\ \arabic{subsection}. }

  \definecolor{darkrosso}{RGB}{100,13,20}
\definecolor{lightor}{RGB}{248,150,30}
\newcommand{\ee}{\end{equation}}
\newcommand{\be}{\begin{equation}}
\newcommand{\bea}{\begin{align}}
\newcommand{\eea}{\end{alig}}

\newcommand   \cO {\mathcal{O}}

\newcommand{\identity}{\mathbbm{1}}

\title{\LARGE\color{rossoCP3}{\textbf{ The $\theta$-angle and axion physics \\ of \\ two-color QCD at fixed baryon charge \\~\\Part I}}}
\author{\normalsize Jahmall \textsc{Bersini}$^{\dagger\blacktriangle}
$, Alessandra \textsc{D'Alise}$^{\star\color{rossoCP3}{\heartsuit}}$, Francesco \textsc{Sannino}$^{\ddagger \color{rossoCP3}{\diamondsuit}}$$^{\clubsuit\color{rossoCP3}{\heartsuit}}$$^\spadesuit$, Matías \textsc{Torres}$^{\sharp\color{rossoCP3}{\heartsuit}}$}

\date{}
\begin{document}
\maketitle 
\let\thefootnote\relax\footnote{$^\dagger$ Electronic address: \textcolor{rossoCP3}{\href{mailto:jbersini@irb.hr}{\textcolor{rossoCP3}{jbersini@irb.hr}}}\\$^\star$ Electronic address: \textcolor{rossoCP3}{\href{mailto:alessandra.dalise@unina.it}{\textcolor{rossoCP3}{alessandra.dalise@unina.it}}}\\$^\ddagger$ Electronic address: \textcolor{rossoCP3}{\href{mailto:sannino@cp3.sdu.dk}{\textcolor{rossoCP3}{sannino@cp3.sdu.dk}}}\\$^\sharp$ Electronic address: \textcolor{rossoCP3}{\href{mailto:matiasignacio.torressandoval@unina.it}{\textcolor{rossoCP3}{matiasignacio.torressandoval@unina.it}}}}\vspace{0.5mm} 
\begin{center}
\footnotesize{$^\blacktriangle$ Rudjer Boskovic Institute, Division of Theoretical Physics, Bijeni\v cka 54, 10000 Zagreb, Croatia }\\
\footnotesize{$\color{rossoCP3}{^\heartsuit}$ Dipartimento di Fisica ``E. Pancini", Università di Napoli Federico II - INFN sezione di Napoli, Complesso Universitario di Monte S. Angelo Edificio 6, via Cintia, 80126 Napoli, Italy}\\
\footnotesize{$^\clubsuit$ Scuola Superiore Meridionale, Largo S. Marcellino, 10, 80138 Napoli NA, Italy}\\
\footnotesize{$\color{rossoCP3}{^\diamondsuit}$ CP$^3$-Origins and D-IAS, Univ. of Southern Denmark, Campusvej 55, 5230 Odense M, Denmark}\\
\footnotesize{$^\spadesuit$ CERN, Theoretical Physics Department, 1211 Geneva 23, Switzerland}
\end{center}
\begin{abstract}
We analyze the impact of the $\theta$-angle and axion dynamics for two-color (in fact any $Sp(2N)$) QCD at nonzero baryon charge and as a function of the number of matter fields on the vacuum properties, the pattern of chiral symmetry breaking as well as the  spectrum of the theory. We show that the vacuum acquires a rich structure when the underlying $CP$ violating topological operator is added to the theory. We discover novel phases and analyse the order of their transitions characterizing the dynamics of the odd and even number of flavours.  We further determine the critical chemical potential as function of the $\theta$ angle separating the normal from the superfluid phase of the theory. 
Our results will guide numerical simulations and novel tests of the model's dynamics. The results are also expected to better inform phenomenological applications of the model ranging from composite Higgs physics to strongly interacting massive dark matter models featuring number changing interactions. In the companion work \cite{PartII} we repurpose and upgrade the approach to determine the impact of the $\theta$-angle and axion physics on non-perturbative near conformal dynamics related to the fixed baryon charge sector.

{\footnotesize  \it Preprint: RBI-ThPhys-2022-31}
\end{abstract}
\newpage
\tableofcontents

\section{Introduction}
\label{intro}

Understanding the dynamics of strongly coupled theories remains an open challenge making it one of the most fascinating topics in theoretical physics. Partial progress has been made both analytically and numerically via a number of tools developed in the past decades. Quantum Chromo Dynamics (QCD),  featuring three colors and several light flavors, has been chosen by Nature as one of the pillars of the Standard Model of particle interactions. QCD therefore stands out within the landscape of strongly coupled theories. However, several relevant questions remain unanswered about its dynamics from the QCD physical spectrum to the interplay between chiral symmetry breaking and confinement, to its phase structure as a function of light matter fields (conformal window), temperature and matter density. For example, knowledge about the phase diagram informs a number of physical applications ranging from the inner structure of Neutron stars \cite{Gandolfi:2019zpj} to the thermal history of the early Universe \cite{Boeckel:2010bey} and last but not least to applications beyond standard model physics and cosmology (see \cite{Sannino:2009za,Cacciapaglia:2020kgq}  and references therein). Another unsolved puzzle is the absence or unexplained strong suppression  \cite{Baker:2006ts,Abel:2020pzs} of the otherwise legitimate presence in the theory of the topological term responsible for strong $CP$ violation. Its possible resolutions in terms of either the axion physics \cite{peccei1977cp,peccei1977constraints,Weinberg:1977ma,Wilczek:1977pj,Kim:1979if,Shifman:1979if,Dine:1981rt,Mohapatra:1978fy} or  its alternatives such as the ones in which $CP$ is broken spontaneously
\cite{Nelson:1983zb,Barr:1984fh,Nelson:1984hg} or models in which the emergent Goldstone boson is not an axion (which would
have a non-zero mass) \cite{Hsu:2004mf} (see also \cite{Wilczek:2016gzx}) are under  active experimental and theoretical investigation. 
 
 Reducing the number of colors from three to two provides a number of theoretical advantages and  increases potential phenomenological applications beyond the QCD template. This is due, in a nutshell, to the symplectic nature of the matter representation that enhances the quantum global symmetries for the light flavors from the $SU(N_f)\times SU(N_f)\times U(1)_B$ to $SU(2N_f)$. The enhancement of the quantum global symmetries (demonstrated on the lattice in \cite{Lewis:2011zb,Hietanen:2014xca}) has far-reaching consequences \cite{Appelquist:1999dq,Ryttov:2008xe}, such as the possibility of the minimal construction of composite Goldstone Higgs theories \cite{Kaplan:1983fs} to the natural introduction of number changing operators (stemming from the Wess-Zumino term) crucial to models in which dark matter genesis occurs within the dark sector itself \cite{Carlson:1992fn,Hochberg:2014dra,Hochberg:2014kqa,Hansen:2015yaa}. When discussing non-zero baryon charge the model allows for well defined lattice simulations because the action remains real, differently from ordinary QCD.  For a review of the various applications of the theory beyond the ones mentioned here we refer to \cite{Sannino:2009za,Cacciapaglia:2020kgq}. Last but not least this model together with ordinary QCD has been one of the most studied theories on the lattice as function of light flavors in the hunt for the predicted lower edge of the conformal window \cite{Sannino:2004qp,Dietrich:2005jn,Dietrich:2006cm,Sannino:2009aw} and its dynamical properties \cite{Karavirta:2011zg,Ohki:2010sr,Rantaharju:2014ila,Lewis:2011zb,Hietanen:2013fya,Hietanen:2014xca,Arthur:2016dir, Leino:2018yfd,Karavirta:2011zg,Bursa:2010xn,Hayakawa:2013yfa,Appelquist:2013pqa,Leino:2016njf,Suorsa:2016jsf}, finite baryon density 
\cite{Nakamura:1984uz,Hands:1999md,Kogut:2001if,Kogut:2002cm,Muroya:2002ry,Hands:2006ve,Cotter:2012mb,Boz:2013rca,Braguta:2016cpw,
Holicki:2017psk,Bornyakov:2017txe,Boz:2018crd,Astrakhantsev:2018uzd,Boz:2019enj,Iida:2019rah,Wilhelm:2019fvp,Bornyakov:2020kyz,Iida:2020emi,Astrakhantsev:2020tdl,Khunjua:2020xws,Kojo:2021knn} and last but not least the investigation of gravity-free asymptotic safety at large number of matter fields (proven rigorously in \cite{Litim:2014uca,Litim:2015iea} for gauge-Yukawa theories) and suggested for gauge-fermion theories in \cite{Antipin:2017ebo} investigated on the lattice in \cite{Leino:2019qwk,Rantaharju:2021iro}.

 \bigskip 
 
 The goal of our work is to go beyond the state-of-the-art by providing an in depth analysis of the $\theta$-angle and axion physics at non-zero baryon chemical potential of two-color QCD. Earlier studies appeared in \cite{Metlitski:2005db}.  We organise our work by introducing in the next Section the two-color effective pion Lagrangian at non-zero baryon charge, including both the $\theta$-angle term as well as the axion field.  In Section~\ref{Vacuum} we will determine the  vacuum structure of the theory both in the normal and superfluid phase as a function of the different number of matter fields. Here we will show how new phases emerge due to the interplay of the $\theta$-angle term, quark masses, and baryon chemical potential. We will also characterise the type of phase transitions. In particular, we will study the dynamics of $N_f = 2,3,6,7,8$  flavors and unveil general properties depending on the even versus odd number of flavors. We stop at eight flavours since around and above this number one expects the dynamics to either become near conformal or conformal \cite{Sannino:2009aw}. The analysis for near conformal dynamics is performed in \cite{PartII}.
Here we discover that in the case of even flavors we observe a first order $CP$ phase transition for $\theta=\pi$ both in the normal and superfluid phase, except for the case of $N_f=2$ for which in the superfluid case the transition disappears as also discussed in \cite{Metlitski:2005db}. For the odd case, the normal phase still supports $CP$ breaking at  $\theta=\pi$ while in the superfluid phase the $\theta=\pi$ is replaced by novel first order phase transitions for $\theta = \frac{\pi}{2}$ and $\theta=\frac{3\pi}{2}$.   We summarize the patterns of chiral symmetry breaking in the normal and superfluid phases and determine the spectrum and associated dispersion relations in Section~\ref{SBP}.  We offer our conclusions in Section~\ref{Conclusions}.

\section{Two-color chiral Lagrangian }
\label{qcdsymmetries}

 The Lagrangian of $N_{f}$  Dirac fermions transforming according to the fundamental representation of two-color QCD reads:
\begin{equation}
        \mathcal{L}=- \frac{1}{4 g^2} {\vec{G}_{\mu \nu}} \cdot   {\vec{G}^{\mu \nu}} + i\bar{\mathcal{Q}}  \bar{\sigma}^{\nu} \left[ \partial_\nu - i \vec{G}_{\nu} \cdot \frac{\vec{\tau}}{2}\right] \mathcal{Q} - \frac{1}{2}m_q \mathcal{Q}^T \tau_2 E \mathcal{Q} + {\rm h.c.} \ .
\end{equation}
 Here ${G}_{\mu \nu}^a$ and ${G}_{\mu}^a$ with $a=1,2,3$ are respectively the gluon field strength and the field itself, $\tau^a$ are the Pauli matrices for the $SU(2)$ color group, and $\mathcal{Q}^{c,i}_{\alpha}$ is a two-spinor fermion field transforming according to the fundamental representation of color with $c=1,2$ and $i=1,\cdots, 2N_f$.  In terms of  $q_{L,R}$, which are the original left and right handed quarks stemming from the Dirac notation it reads
   \begin{equation}
 \mathcal{Q}= \begin{pmatrix}
    q_L\\
    i \sigma_2 \tau_2 q^{\ast}_R 
    \end{pmatrix} \ .
 \end{equation}
At zero fermion mass (i.e. $m_q = 0$)  the theory exhibits the classical $U(2N_f)$ symmetry  broken at the quantum level by the Adler-Bell-Jackiw anomaly to $SU(2N_f)$.  The Dirac mass term breaks explicitly the symmetry to $Sp(2N_f)$ and the $2N_f\times 2N_f$ matrix $E$ reads: 
  \begin{equation}
    E=\begin{pmatrix}
    0 & 1\\
    -1 & 0
    \end{pmatrix}\otimes\mathbbm{1}_{N_f}\ .
\end{equation}
At small number of flavors the dynamics is expected to be strong enough for a fermion condensate to form breaking the $SU(2N_f)$ global symmetry spontaneously to a smaller subgroup expected to be the maximal diagonal one. However, only recently, first-principle lattice simulations have been able to demonstrate this pattern of chiral symmetry breaking in \cite{Lewis:2011zb} which has been further confirmed in \cite{Hietanen:2014xca} for two Dirac flavors.   Increasing the number of flavors the dynamics can change and one expects the existence of a critical number of flavors $N^\ast_{f}$ above which an IR interacting conformal fixed point emerges. This dynamics has been investigated theoretically \cite{Sannino:2009aw} and via first principle lattice simulations \cite{Amato:2018nvj}. A recent up-to-date review can be found in \cite{Cacciapaglia:2020kgq}.   In fact, even the dynamics at very large number of flavours is extremely interesting both theoretically and phenomenologically. It is also being investigated on the lattice \cite{Leino:2019qwk} searching for the existence of an interacting non-perturbative UV fixed point \cite{Antipin:2017ebo} similar to the one shown to exist in \cite{Litim:2014uca,Litim:2015iea} for gauge-Yukawa theories. 
For any $2<N_f < N^\ast_{f}$ it is therefore natural to expect the pattern of spontaneous symmetry breaking to be $SU(2N_f) \rightarrow Sp(2N_f)$. Generalizing this theory to arbitrary $N$ while keeping the same global symmetries requires considering $Sp(2N)$ gauge groups and the associated conformal window has been discussed in \cite{Sannino:2009aw}.  
The associated chiral Lagrangian embodying the previous pattern of symmetry breaking reads  \cite{Appelquist:1999dq,Sannino:2009za} : 

\begin{equation}
\label{eq:lkinandlmass}
  \mathcal{L}_{\text{eff}}= \nu^2 Tr\{ \partial_\mu\Sigma\partial^\mu\Sigma^\dagger\}+m^2_\pi \nu^2 Tr\{M\Sigma+M^\dagger\Sigma^\dagger\}\ , 
\end{equation}
with $\Sigma$ transforming linearly under a chiral rotation as 
\begin{equation}
\Sigma \rightarrow u \Sigma u^T \ , \qquad u \in SU(2N_f) \ ,
\end{equation}
and the democratic mass matrix $M$ being
\begin{equation}
\label{Massa}
M = - i\sigma_2 \otimes \mathbbm{1}_{N_f}=  \begin{pmatrix}
    0 & -1\\
    1 & 0
    \end{pmatrix}\otimes\mathbbm{1}_{N_f}\ .
\end{equation}

\subsection{Baryon charge}
 
We can take into account a non-vanishing baryon charge by coupling it to a chemical potential $\mu$. The latter can be introduced as the zero component of a background gauge field via the covariant derivative 
\begin{equation} \label{covar}
    \partial_\mu\mapsto D_\mu=\partial_\mu-i\mu\delta_\mu^{0} B  \,, \qquad \quad B \equiv \begin{pmatrix}
1/2 & 0\\
0 & -1/2
\end{pmatrix}\otimes\mathbbm{1}_{N_f}
\end{equation}
which reproduces the usual coupling of the chemical potential to the Noether charge in the Dirac notation. In fact
\begin{equation}
    \Bar{q}\gamma^0 q=\begin{pmatrix}
    q^\ast_L\\
    q^\ast_R
\end{pmatrix}^T \begin{pmatrix}
1 & 0\\
0 & 1
\end{pmatrix}\otimes\mathbbm{1}_{N_f}\begin{pmatrix}
    q_L\\
    q_R
\end{pmatrix}=\mathcal{Q}^\dagger \underbrace{\begin{pmatrix}
1 & 0\\
0 & -1
\end{pmatrix}\otimes\mathbbm{1}_{N_f}}_{\equiv 2 B}\mathcal{Q}=2 \mathcal{Q}^\dagger B\mathcal{Q}\ .
\end{equation}
Notice that the baryon charge generator belongs to the $su(2N_f)$ algebra being proportional to the third Pauli matrix.  


From the form of the $B$ matrix, we see that for non-zero $\mu$ the Lagrangian is no longer invariant under $SU(2N_f)$ transformations.
To fix the ideas, when the mass term is zero the $SU(2N_f)$ symmetry is explicitly broken as 
\begin{equation}
    SU(2N_f)\xrightarrow{m=0, ~ \mu\neq 0} SU(N_f)_L\times SU(N_f)_R\times U(1)_B\ ,
\end{equation}
conversely, for $m \neq 0$ one has
\begin{equation}
    SU(2N_f)\xrightarrow{m\not=0,~ \mu\neq 0} SU(N_f)_V\times U(1)_B\ .
\end{equation}
By using the covariant derivative \eqref{covar} into eq.\eqref{eq:lkinandlmass}, we obtain the effective Lagrangian describing the theory at non-zero baryon charge

\begin{equation}
\label{eq:leffective}
\begin{split}
    \mathcal{L}_{B}&= \nu^2 Tr\{ \partial_\mu\Sigma\partial^\mu\Sigma^\dagger\}+4\mu \nu^2 Tr\{B\Sigma^\dagger\partial_0\Sigma\}+m^2_\pi \nu^2 Tr\{M\Sigma+M^\dagger\Sigma^\dagger\} \\ 
    & +2\mu^2 \nu^2\left[Tr\{\Sigma B^T\Sigma^\dagger B\}+Tr\{BB\}\right]\ .
\end{split}
\end{equation}

 \subsection{The $\theta$-angle and the $\mathrm{U(1)}$ problem}
 \label{includingthethetaangle}
In order to discuss the physics of the $\theta$-angle and of the axial anomaly \cite{witten1979current,veneziano1979u,crewther1979chiral}, we introduce the  topological charge density    
\begin{equation}
\label{qxdef}
     q(x)=\frac{g^2}{64\pi^2} \epsilon^{\mu\nu\rho\sigma}F^a_{\mu\nu}F^a_{\rho\sigma}\ ,
\end{equation}
which we incorporate in the effective Lagrangian as
 \be
 \label{ourthetalag}
\mathcal{L}_{q(x)}=\frac{i}{4}q(x)\Tr\{\log \Sigma-\log \Sigma^\dagger\}-\theta q(x)+\frac{q(x)^2}{4 a \nu^2}\ .
\ee
Here  the new $\Sigma$ is related to the old one that transforms only under  $SU(2N_f)$  via \cite{DiVecchia:2013swa}
\begin{equation}
    \Sigma \rightarrow     \Sigma   e^{i \frac{S}{\sqrt{2N_f}} \mathbbm{1}_{2N_f}}\, , 
\end{equation}
with the $S$-field a singlet of $SU(2N_f)$  transforming under the anomalous $U(1)_A$, and it is parent to the $\eta^\prime$ particle in ordinary QCD.\\

The minimal choice to neglect orders higher than $q^2(x)$ is justified at large number of colors, see discussion in \cite{DiVecchia:2013swa}. Here, to keep the same pattern of chiral symmetry breaking the $SU(2)$ of color generalizes to $Sp(2N)$ and not $SU(N)$.  The linear term allows accommodating for the $U(1)_A$ anomaly. The coefficient of the quadratic term in $q(x)$ is known as the topological susceptibility of the Yang-Mills theory. The coefficients of these terms are coherently chosen such that we reproduce the axial anomaly $\partial_\mu J^\mu_5=4N_fq(x)$ in two-color QCD with quarks in the fundamental representation \cite{di2014physics}. Notice that $q(x)$ is a background auxiliary field introduced to implement the anomalous transformation at the action level in linear fashion. It can, therefore,be integrated out via its equations of motion, yielding the following effective Lagrangian

\begin{align}
\label{lagtheta}
    \mathcal{L}_{\theta}&= \nu^2 Tr\{ \partial_\mu\Sigma\partial^\mu\Sigma^\dagger\}+4\mu \nu^2 Tr\{B\Sigma^\dagger\partial_0\Sigma\}+m^2_\pi \nu^2 Tr\{M\Sigma+M^\dagger\Sigma^\dagger\} \nonumber \\ &+2\mu^2 \nu^2\left[Tr\{\Sigma B^T\Sigma^\dagger B\}+Tr\{BB\}\right]  -a \nu^2\left(\theta-\frac{i}{4}Tr\{\log \Sigma - \log \Sigma^\dagger \}\right)^2 \,.
\end{align}
The action of the symmetry groups is summarised in \autoref{BBB} where the axial generator $A$ in our notation is
\begin{equation}
    A=\frac{1}{2}\begin{pmatrix}
    1 & 0\\
    0 & 1\\
    \end{pmatrix}\otimes\mathbbm{1}_{N_f}\ .
\end{equation}
\begin{table}[h!]  
\centering
\begin{tabular}{l|l|lllllll}
               & $[ SU(2) ]$ & $SU(N_f)_L$                   & $\times$ & $SU(N_f)_R$                   & $\times$ & $U(1)_V$ & $\times$ & \multicolumn{1}{l|}{$U(1)_A$} \\ \hline
$q_L$       & $\Box$                 & $\Box$                        &          & \textbf{1}                             &          & +1       &          & \multicolumn{1}{l|}{+1}       \\
$i\sigma_2\tau_2 q^\ast_R$ & $\Box$                 & \textbf{1}                             &          & $\Bar{\Box}$                  &          & $-$1       &          & \multicolumn{1}{l|}{+1}       \\ \hline \\  
               & $[ SU(2) ]$ & $SU(2N_f)$                    & $\times$ & \multicolumn{1}{l|}{$U(1)_A$} &          &          &          &                               \\ \cline{1-5}
$\mathcal{Q}$         & $\Box$                 & {\color[HTML]{000000} $\Box$} &          & \multicolumn{1}{l|}{+1}       &          &          &          &                               \\ \cline{1-5}
\end{tabular}

\caption{Transformation properties of $q_L$, $i\sigma_2\tau_2 q^\ast_R$ and $\mathcal{Q}$ under the action of the symmetry groups.} \label{BBB}
\end{table}\\

 \subsection{The axion and the baryon charge}
 \label{includingtheaxion}

For the Standard Model three colors QCD there is no experimental evidence of strong $CP$ violation. This constrains the associated value of the effective theta angle  $\bar{\theta} = \theta + {\rm arg}\det(M)$, with $M$ the physical quark mass matrix, to be $\bar{\theta}< 10^{-10}$. The limit comes from the bound on the neutron electric dipole moment $|d_n| = C_{\rm EDM} e ~\bar{\theta}<1.8\times 10^{-26} e~{\rm cm}$ \cite{Baker:2006ts,Abel:2020pzs}, where $ C_{\rm EDM} = 2.4(1.0)\times 10^{-16}~{\rm cm}$ \cite{Pospelov:1999mv}  is related to the effective nucleon interactions with the axion explained in the Appendix of \cite{DiVecchia:2013swa}.  From a theoretical  viewpoint, a tiny value of a physical parameter unprotected by any symmetry requires an explanation.   One solution for the so-called strong $CP$ problem was proposed in the $70$s by R. Peccei and H. Quinn  \cite{peccei1977cp,peccei1977constraints}. The proposal makes use of an additional $U(1)_{PQ}$ symmetry which is quantum mechanically anomalous and spontaneously broken leading to the axion as an extra (pseudo-)Goldstone boson. Axion physics has led to a great deal of both theoretical and phenomenological work including active experimental searches beyond the original QCD axion \cite{Essig:2013lka}. An alternative solution to the strong $CP$ problem featuring new composite dynamics and heavier (if not completely absent) axions was proposed in \cite{Hsu:2004mf} and reconsidered in \cite{Wilczek:2016gzx} and applied  extensively in model building in \cite{Gaillard:2018xgk,Gavela:2018paw,DiLuzio:2021pxd}. 

Of course, when discussing the $\theta$-angle physics here, depending on the physical application of the model, we can allow for new sources of $CP$ breaking, perhaps relevant for some models of dark matter or Standard Model secluded sectors \cite{Essig:2013lka}. Nevertheless, it is interesting to entertain the possibility that an axion is present in the theory and therefore explore its finite chemical potential dynamics and impact. We denote by $\nu_{PQ}$ the scale of $U(1)_{PQ}$ spontaneous symmetry breaking and by $a_{PQ}$ the coefficient of the $U(1)_{PQ}$ anomalous term. We include the axion in our theory by extending \eqref{lagtheta} to the following effective Lagrangian 
\begin{equation}
\small
\begin{split}
   \mathcal{L}_{\hat{a}}&= \nu^2 Tr\{ \partial_\mu\Sigma\partial^\mu\Sigma^\dagger\}+\nu_{PQ}^2 \partial_\mu N\partial^\mu N^\dagger+4\mu \nu^2 Tr\{B\Sigma^\dagger\partial_0\Sigma\}+m^2_\pi \nu^2 Tr\{M\Sigma+M^\dagger\Sigma^\dagger\}\\& +2\mu^2 \nu^2\left[Tr\{\Sigma B^T\Sigma^\dagger B\}+Tr\{BB\}\right] -a \nu^2\left(\theta-\frac{i}{4}Tr\{\log \Sigma - \log \Sigma^\dagger \}  -\frac{i}{4}a_{PQ}(\log N - \log N^\dagger) \right)^2\ .
\end{split}
\end{equation}
The details on how to construct the extended effective theory can be found in \cite{DiVecchia:2013swa}.
 
\section{Determining the Vacuum}
\label{Vacuum}

In this section, we focus on the vacuum structure of theory \eqref{lagtheta} in the presence of the $\theta$-angle and finite baryon charge. In particular, we will first study the general conditions determining the classical vacuum solution as a function of $\theta$ and $\mu$. Armed with these results, we then carefully analyze the properties of the vacuum in the concrete cases $N_f=2\,,3\,,6\,,7\,,8$ and $N_f$ arbitrary. To begin with, we notice that for vanishing $\theta$ the vacuum is determined by the competition of mass and baryon chemical potential. In other words, the ground state can be written as \cite{kogut2000qcd}
\begin{equation}
\label{eq:ansatsvacuum}
    \Sigma_c=\begin{pmatrix}
    0 & \mathbbm{1}_{N_f}\\
    -\mathbbm{1}_{N_f} & 0
    \end{pmatrix}
\cos\varphi+i \begin{pmatrix}
    \mathcal{I} & 0\\
    0 & \mathcal{I}
    \end{pmatrix}\sin\varphi \qq{where} \mathcal{I}=\begin{pmatrix}
    0 & -\mathbbm{1}_{N_f/2}\\
    \mathbbm{1}_{N_f/2} & 0
    \end{pmatrix}\ ,
\end{equation}
where the angle $\varphi$ is determined by the equations of motion.
 
To study the effect of $\theta$ on the vacuum solutions it is convenient to introduce the Witten variables $\alpha_i$ as \cite{Witten:1980sp} 
\be \label{anscomp}
\Sigma_0 = U(\alpha_i) \Sigma_c \,, \qquad U(\alpha_i) \equiv \text{diag}\{e^{-i \alpha_1}\,, \dots \,,e^{-i \alpha_{N_f}},e^{-i \alpha_1}\,, \dots \,,e^{-i \alpha_{N_f}} \} \,.
\ee
 These variables reflect the possibility to generalize the mass matrix in \eqref{Massa} by replacing the $\identity_{N_f}$ with \\$ \text{diag}\{e^{-i \alpha_1}, \dots,e^{-i \alpha_{N_f}}\}$. Each phase is the overall axial transformation for each left-right quark pair. The reason why we cannot consider a single (odd number of) Weyl fermion(s) at the time (as is the case for real gauge representations) is that one cannot write a mass term in this case and the theory overall suffers from a topological anomaly  \cite{Witten:1982fp}.

We take the above to be the general ansatz for the vacuum. The Lagrangian evaluated on this ansatz reads
\begin{equation}
\label{NCTlag}
   \mathcal{L}_{\theta}[\Sigma_0]=\nu ^2\left[4  m_\pi^2  X  \cos \varphi +2 \mu ^2  N_f \sin ^2\varphi -a  \bar{\theta}^2 \right]
\end{equation}
where for later convenience we introduced 
\be
\bar{\theta}=\theta -\sum_i^{N_f} \alpha_i \,, \qquad X=\sum_i^{N_f} \cos \alpha_i 
\ee
where $\bar{\theta}$ is the effective theta angle that enters physical observables.
The equations of motion read
\begin{align} \label{eqalphanoconf1}
   \sin \varphi \left(  N_f \cos \varphi - \frac{m_\pi^2}{\mu ^2}  X \right) &=0  \\ \label{eqalphanoconf2}
   2 m_\pi^2  \sin \alpha_i  \cos \varphi&= a \bar{\theta} \,, \quad i=1, .., N_f
\end{align}
and the energy of the system is 
\begin{align}
\label{en1senzadil}
  E &=-\nu ^2 \left[4 m_{\pi }^2 X -a \bar{\theta}^2\right] \,, \quad &\text{normal phase} \ &  \left(\varphi=0\right) \\ \label{en2senzadil}
    E&=-\nu ^2\left[2\frac{ N_f^2 \mu
   ^4+  m_{\pi }^4 X^2}{N_ f\mu ^2} -a \bar{\theta}^2\right]\,, \quad &\text{superfluid phase} \   &\left(\cos\varphi=\frac{m_\pi^2}{ N_f \mu ^2}  X\right)\ .
\end{align}
Note that when $a\gg m_\pi$ all the $\theta$-dependence is contained in an effective pion mass $m_\pi^2(\theta) \equiv \frac{m_\pi^2 X}{N_f} $.
For $\theta=0$ (i.e. $U(\alpha_i)=\identity$) one has $X=N_f$ and $\bar{\theta}=0$, and we recover the results in \cite{kogut2000qcd}. In this case, a normal to superfluid phase transition occurs when $\mu = m_\pi$, i.e. when $\cos\varphi = 1$. 
For $\theta \neq 0$ the $\theta$-dependence of the energy may be different in the two phases. Therefore, to find the conditions for the onset of the superfluid phase we first need to determine the $\theta$ vacuum in both phases. To this end, we observe that in the normal (superfluid) phase the energy is minimized when $X$ ($X^2$) is maximized. 
In the former case, the Witten variables are related to $\theta$ by the well-known equation
\begin{align}
2 m_\pi^2 \sin \alpha_i &= a \bar{\theta} =  a \left(\theta-\sum_i^{N_f} \alpha_i \right) \ .\label{thetaeq}
\end{align} 
For the general solution we must have for any $\bar{\theta}$ fixed  $\sin \alpha_i = \sin \alpha_j$. To solve for the $\alpha_i$  we consider the expansion in the parameter $\frac{m_\pi^2}{a}$ that we take to be small. Concretely, at the leading order one needs to solve for $\bar{\theta}=0$  and the angles $ \alpha_{i}$ satisfy
\begin{align} \label{Solgen2}
    \alpha_{i}=\begin{cases}
  \pi-  \alpha ,\qquad  & i=1,\dots,n \\
  \alpha ,\qquad & i=n+1,\dots,N_f
    \end{cases} 
\end{align}
where $\alpha$ is the solution of the following modular equation
\begin{align} \label{mods}
n (\pi-\alpha)+(N_{f}-n) \alpha=\theta \ \text{Mod }2 \pi \ .
\end{align}
The modulo comes from the fact that if a solution $\{\alpha_i\}$ of eq.\eqref{thetaeq} is found, then it is possible to build another solution as follows
\be
\alpha_1(\theta+2\pi)=\alpha_1(\theta)+ 2 \pi \,, \qquad \quad \alpha_i(\theta+2\pi)=\alpha_i(\theta) \,, \quad i=2, \dots , N_f \,.
\ee
However, since the physics depends only on $e^{-i \alpha_i}$, the dynamics is invariant under $\theta \to \theta +2 \pi$.
The solution of eq.\eqref{mods} can be written as
\begin{equation} \label{Solgen}
    \alpha=\frac{\theta+ (2k-n)\pi}{(N_{f}-2n)},\quad k=0,\dots, N_f-2n-1 \,, \quad n=
    0,..., \left[\frac{N_f-1}{2}\right] \,.
\end{equation}
The range for $k$ above emerges because for $k\geq N_f-2n$ we repeat the solution for a given $n$. Note that when $n \neq 0$, the vacuum spontaneously breaks $Sp(2 N_f)$ because of the different phases for each quark flavour. 


It is well-known that $CP$ is preserved when $\bar{\theta}=0$. For equal mass quarks as considered here, this happens when $m_\pi=0$ or $\theta=0$. On the other hand, for $\theta=\pi$ the Lagrangian \eqref{lagtheta} possess $CP$
symmetry but in the normal phase the latter is spontaneously broken by the vacuum \cite{Dashen:1970et,DiVecchia:2013swa,Gaiotto:2017tne,DiVecchia:2017xpu} \footnote{See also the discussion about the impact on the  $\theta=\pi$ solution due to quark masses ordering provided in the appendix of \cite{DiVecchia:2013swa}. }, leading to a strong $\theta$-dependence near $\theta=\pi$. In fact, assuming that the ground state does not break $Sp(2 N_f)$ spontaneously (i.e. $n=0$), the vacua lie at \cite{Gaiotto:2017tne}
\be
U(\alpha_i)=e^{i\frac{\theta+2 \pi k}{N_f}}\identity_{2N_f} \ .
\ee
For $\theta=\pi$ one has $X=\cos\left( \frac{(2k+1)\pi}{N_f}\right)$, which is maximized when $k=0$ and $k=N_f-1$, that is
\be \label{2fold}
 U(\alpha_i)=e^{\frac{i \pi}{N_f}}\identity_{2N_f}  \,, \qquad  U(\alpha_i)=e^{- \frac{i \pi}{N_f}}\identity_{2N_f} \ .
\ee
The two solutions are related by a $CP$ transformation $U \to U^\dagger$ and thus $CP$ is spontaneously broken. For $N_f > 2$ the minima are separated by an energy barrier while for $N_f=2$ the leading order quark-mass induced potential vanishes\footnote{As well explained in the work by Smilga \cite{Smilga:1998dh} this phenomenon occurs because the trace of $SU(2)$ matrices are real. }, apparently leading to a paradoxical situation according to which one has massless pions and no explicit breaking of chiral symmetry. The paradox is simply resolved by going to  higher orders in the mass for the chiral Lagrangian. Once these corrections are taken into account  they lift, as expected, the  vacuum degeneracy yielding two minima separated by a barrier \cite{Smilga:1998dh, Tytgat:1999yx}.
The spontaneous breaking of $CP$ at $\theta=\pi$ is known as Dashen’s phenomenon \cite{Dashen:1970et} and has been thoroughly studied in the literature \cite{Witten:1980sp, Creutz:1995wf, Smilga:1998dh, Tytgat:1999yx, Gaiotto:2017tne, DiVecchia:2017xpu, Kitano:2021jho}. As we shall see, in the superfluid phase $CP$ may be violated or not at $\theta=\pi$ depending on the value of $N_f$. Once determined the ${\alpha_i}$ to the leading order in $\frac{a}{m_\pi^2}$, subleading  corrections can be more easily computed. Note that in the superfluid phase the equation of motion \eqref{eqalphanoconf2} becomes 
\be \label{vensup}
 \frac{2 m_\pi^4}{ N_f \mu ^2} X \sin \alpha_i  = a \bar{\theta} \,, \qquad i=1, .., N_f.
\ee
Hence, in this case the natural expansion parameter is $\frac{ m_\pi^4}{a \mu ^2}$ leading to an enhanced suppression  compared to the normal phase. 

\bigskip
We now consider specific values of $N_f$ to cover most of the phase diagram for which chiral symmetry is expected to break spontaneously. In the companion paper \cite{PartII} we will re-examine the values of $N_f$ that are expected to cover the near-conformal dynamics.

\subsection{$\mathrm{N_f = 2}$}
We now focus on the case $N_f=2$ which has been previously considered in \cite{Metlitski:2005db} at the leading order in $\frac{m_\pi^2}{a}$. In this limit, the two angles $\{\alpha_{1},\alpha_{2}\}$ satisfy $\alpha_{1}+\alpha_{2}=\theta + 2k\pi$. Additionally, because $\sin \alpha_{1} = \sin \alpha_{2}$ we have
\begin{align}
    \sin\alpha_{1}=\sin\left( \theta+2k\pi-\alpha_{1}\right)
\end{align}
yielding the solutions 
\begin{align}
    \sin\alpha_{1}=\sin\alpha_{2}=
 {\sin \left( \frac{\theta}{2} + k\pi  \right)}. 
\end{align}
 Only $k=0$ and $k=1$ are independent solutions of the equations of motion and agree with the general result in equation \eqref{Solgen}  for the $\alpha_{i}$ for $N_f=2$  that has the two solutions corresponding to $n=0$ and $k=0,1$ (i.e. $\{\alpha_1, \alpha_2 \} = \{\frac{\theta}{2}, \frac{\theta}{2}\}$ and $\{\alpha_1, \alpha_2 \}= \{\frac{\theta+2\pi}{2}, \frac{\theta+2\pi}{2} \} $).

\begin{figure}[h!]
    \centering
    \subfloat[\centering $\theta$-dependence of the energy in the normal phase for $N_f=2$.]{{\includegraphics[scale=0.7]{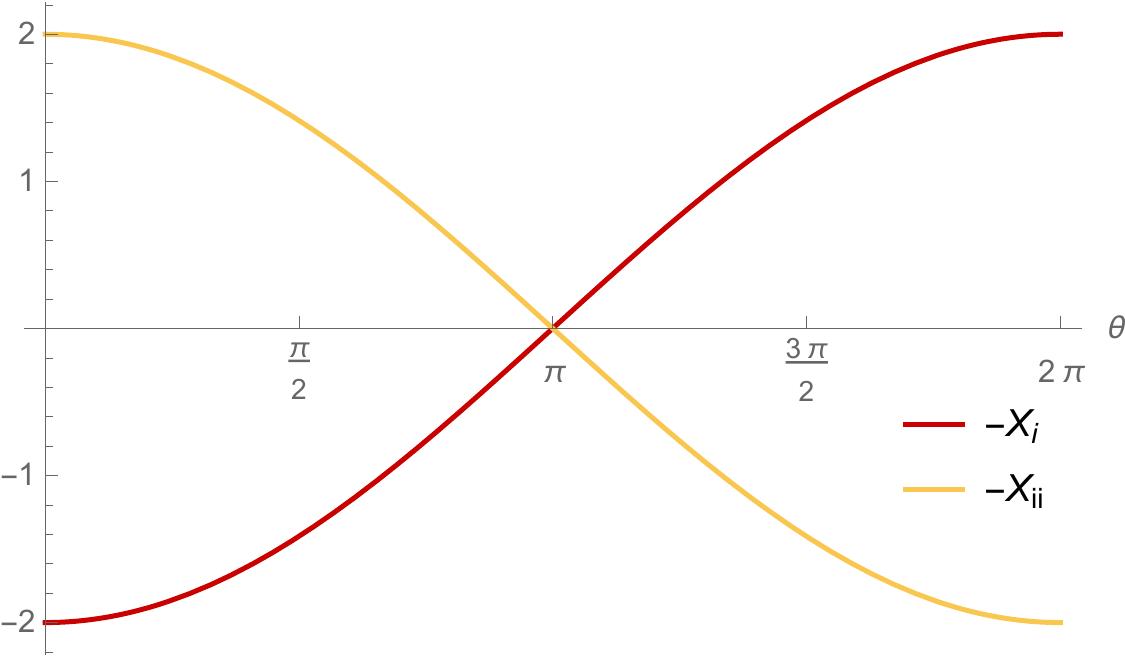} }}%
    \qquad
    \subfloat[\centering $\theta$-dependence of the energy in the superfluid phase for $N_f=2$.]{{\includegraphics[scale=0.7]{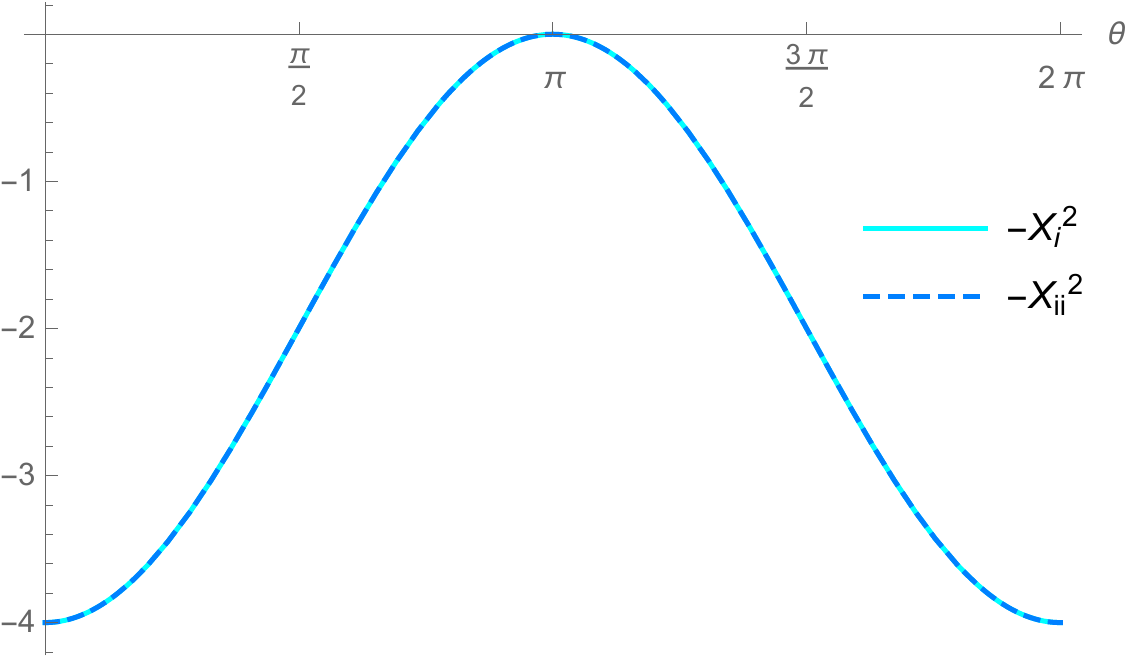} }}%
    \caption{$\theta$-dependence of the energy for $N_f=2$.}%
    \label{fig:energyN2}%
\end{figure}

 \vskip .5cm
 \noindent
 To determine the solution corresponding to the ground state for any value of $\theta$ we need to consider the ground state energy. 
   In the normal phase, the energy is linear in $X$ and therefore the original equation of motion solutions cross at $\theta = \pi$ where Dashen's phenomenon occurs stemming from spontaneous $CP$ symmetry breaking.  Because in the superfluid phase we have that the ground state energy is proportional to $X^2$ the two solutions are identical yielding a degenerate ground state. Interestingly this degeneracy is not lifted by higher order corrections in $\frac{m_\pi^2}{a}$. The $\theta$-dependence in the two phases is shown in Fig.\ref{fig:energyN2}.

 Additionally, when $\theta=\pi$ the effective mass $m_{\pi}^2(\theta) \sim m_\pi^2 \abs{ \cos\left(\frac{\theta}{2}\right)}$ vanishes up to correction of order $\cO\left(\frac{m_\pi^2}{a}\right)$. Therefore, the mass term disappears from the Lagrangian and the global flavor symmetry is again $SU(4)$ consequently leading to massless Goldstones when it spontaneously breaks to $Sp(4)$ \cite{Metlitski:2005db}. Nevertheless, there is no chiral symmetry restoration in the fundamental Lagrangian. As mentioned, this apparent paradox is solved by realising that $SU(4)$ is still broken by higher order mass terms in the effective Lagrangian also for $a \to \infty$ \cite{Smilga:1998dh,Metlitski:2005db,Gaiotto:2017tne}.  
By including the first two subleading corrections in $\frac{m_\pi^2}{a}$, the energy in the two phases reads
\begin{align}
E(\theta)&= -8m_{\pi}^2\nu^{2}\left(\left| \cos \frac{\theta }{2}\right|  +\frac{1}{2} \frac{m_{\pi}^2}{a} \sin ^2\frac{\theta }{2}-\frac{1}{4} \frac{m_{\pi}^4}{a^2} \left| \sin \frac{\theta }{2} \sin \theta \right|\right),\quad \text{
normal phase}\\
 E(\theta)&=- \nu ^2\left(\frac{4 \left(m_{\pi}^4  \cos^2 \frac{\theta }{2} +\mu ^4\right)}{\mu ^2}+\frac{m_{\pi}^8 \sin ^2\theta }{a \mu ^4}-\frac{m_{\pi}^{12} \sin ^2 \theta  \cos \theta}{a^2 \mu ^6}\right),\quad \text{superfluid phase}\ .
\end{align}

At fixed quark masses, the superfluid phase transition occurs at a critical value of the chemical potential given by
\be
\mu_c = m_\pi(\theta) = m_\pi \left[\sqrt{\abs{\cos \frac{\theta}{2}}}+ \mathcal{O}\left(\frac{m_\pi^2}{a} \right)  \right] \,,
\ee
implying that it can be realized for tiny values of the chemical potential when $\theta \sim \pi$. To estimate $\mu_c$ in the region $\theta \sim \pi$, we introduce $\phi \equiv \theta - \pi$ and take into account the leading correction in $\frac{m_\pi^2}{a}$. As a result, we have that for small $\abs{\phi}$ the critical chemical potential in the region $\theta \sim \pi$ reads
\be
\mu_c \sim m_\pi  \sqrt{\frac{m_\pi^2}{a}+\frac{\abs{\phi}}{2}}  \,,
\ee
and vanishes for $a \to \infty$ and $\phi \to 0$.

\begin{figure}[h!]
    \centering
    \subfloat[\centering $CP$ order parameter for $N_f=2$ in the normal phase.]{{\includegraphics[scale=0.7]{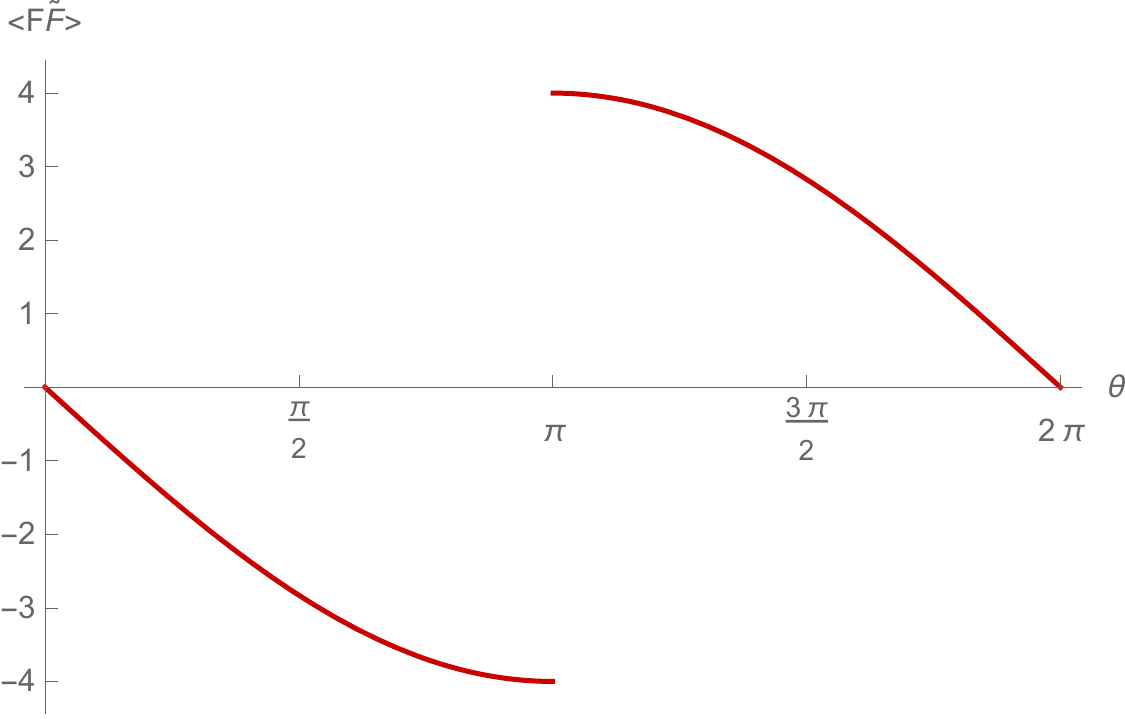} }}%
    \quad
    \subfloat[\centering$CP$ order parameter for $N_f=2$ in the superfluid phase.]{{\includegraphics[scale=0.7]{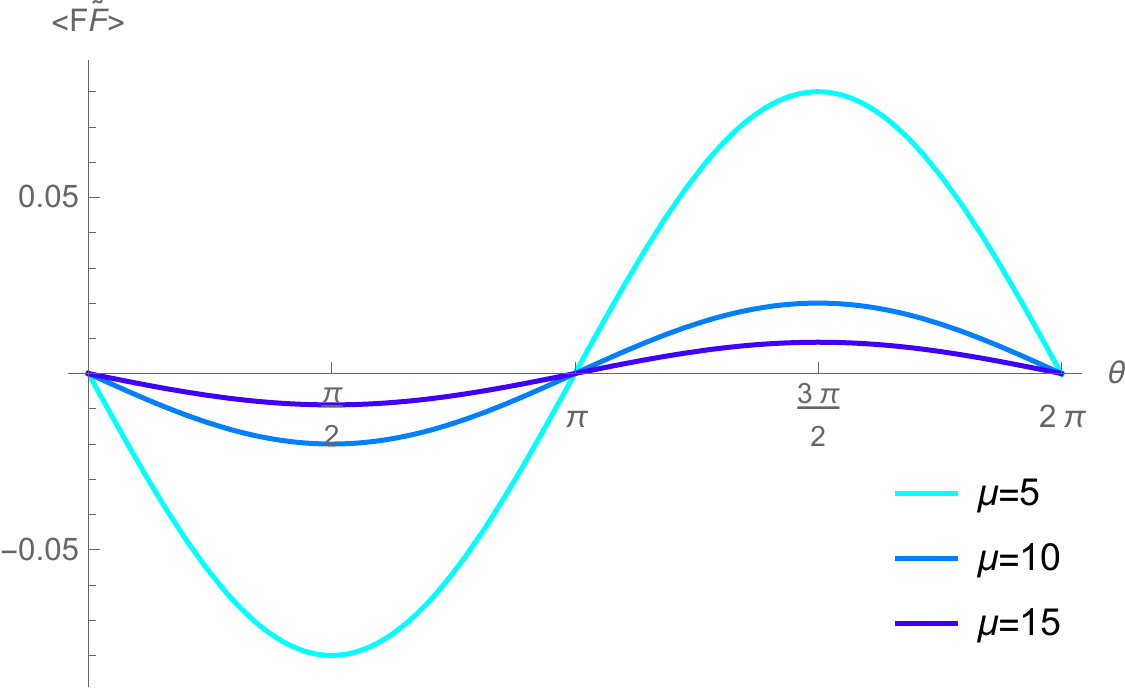} }}%
    \caption{$CP$ order parameter as a function of $\theta$ both in the normal and superfluid phase for $N_f=2$.}%
    \label{fig:CPNF2}%
\end{figure}

Still, in the superfluid phase the energy is an analytic function of $\theta$. This can be better appreciated by analysing the $CP$ order parameter $\expval{F \Tilde{F}} \propto -\pdv{E}{\theta}$ shown in Fig. \ref{fig:CPNF2} for the two phases. The normal phase is characterised by a discontinuous $CP$ order parameter at $\theta=\pi$ while it is clear from the right hand panel of the figure that the superfluid phase displays a smooth behaviour for any value of the $\theta$-angle and vanishes for $\theta=\pi$. 
The discontinuity of the $CP$ order parameter in the normal phase translates into a divergent topological susceptibility $\displaystyle{\chi =\frac{\ \partial^2 E}{\partial \theta^2}}$ at $\theta=\pi$. Note that the presence of the baryon chemical potential suppresses $CP$ violation for every value of $\theta$ as can be seen from the right panel of Fig. \ref{fig:CPNF2}.
Finally, we study the effective $\bar{\theta}$ angle for $\theta=\pi$ at higher orders in $\frac{m_\pi^2}{a}$ for the solutions of the equations of motion. We find 
\begin{align}
 \bar{\theta} &= \frac{2 m_\pi^2}{a} \sin \frac{\theta}{2} \underset{\theta=\pi}{=} \frac{2 m_\pi^2}{a} + \cO\left(\frac{m_\pi^6}{a^3} \right) \,, \qquad &\text{normal phase}  \\
  \bar{\theta} &= \frac{m_\pi^4}{a \mu^2} \sin \theta  \underset{\theta=\pi}{=} 0  \,, \qquad & \text{superfluid phase}\ .
  \label{superfase}
\end{align}
Note that eq.\eqref{superfase} is exact to all orders in $\frac{ m_\pi^4}{a \mu ^2}$ since the equation of motion \eqref{vensup} becomes
\be
\frac{m_\pi^4}{a \mu ^2}\sin (2 \alpha ) = \pi-2 \alpha \,,
\ee
and has the solution $\alpha=\pi/2$ (equivalent to $\alpha=3\pi/2$ under a shift $\theta \to \theta + 2\pi$).
Therefore to the present order in the chiral expansion, there is no spontaneous  $CP$  symmetry breaking in the superfluid phase for $\theta=\pi$. 

\subsection{$\mathrm{N_f = 3}$}

Compared to the  previous subsection, the $N_f=3$ case leads to a richer vacuum structure as has been first pointed out in \cite{Smilga:1998dh}, who studied the physics in the normal phase. In fact, we find the following four solutions for the set $\{\alpha_i\}=\{\alpha_1\,, \alpha_2\,,\alpha_3\}$

\be \label{3sol}
\small
  \mathbf{i.} \Bigg\{ \frac{\theta}{3}, \frac{\theta}{3}, \frac{\theta}{3}  \Bigg\} ,  \quad \mathbf{ii.} \Bigg\{ \frac{\theta+2\pi}{3}, \frac{\theta+2\pi}{3}, \frac{\theta+2\pi}{3}  \Bigg\} ,  \quad \mathbf{iii.} \Bigg\{ \frac{\theta+4\pi}{3}, \frac{\theta+4\pi}{3}, \frac{\theta+4\pi}{3}  \Bigg\} ,  \quad \mathbf{iv.} \big\{ \theta-\pi\,, \theta-\pi \,, 2\pi-\theta  \big\}\ .
\ee

\begin{figure}[h!]
    \centering
    \subfloat[\centering $\theta$-dependence of the energy in the normal phase for $N_f=3$.]{{\includegraphics[scale=0.58]{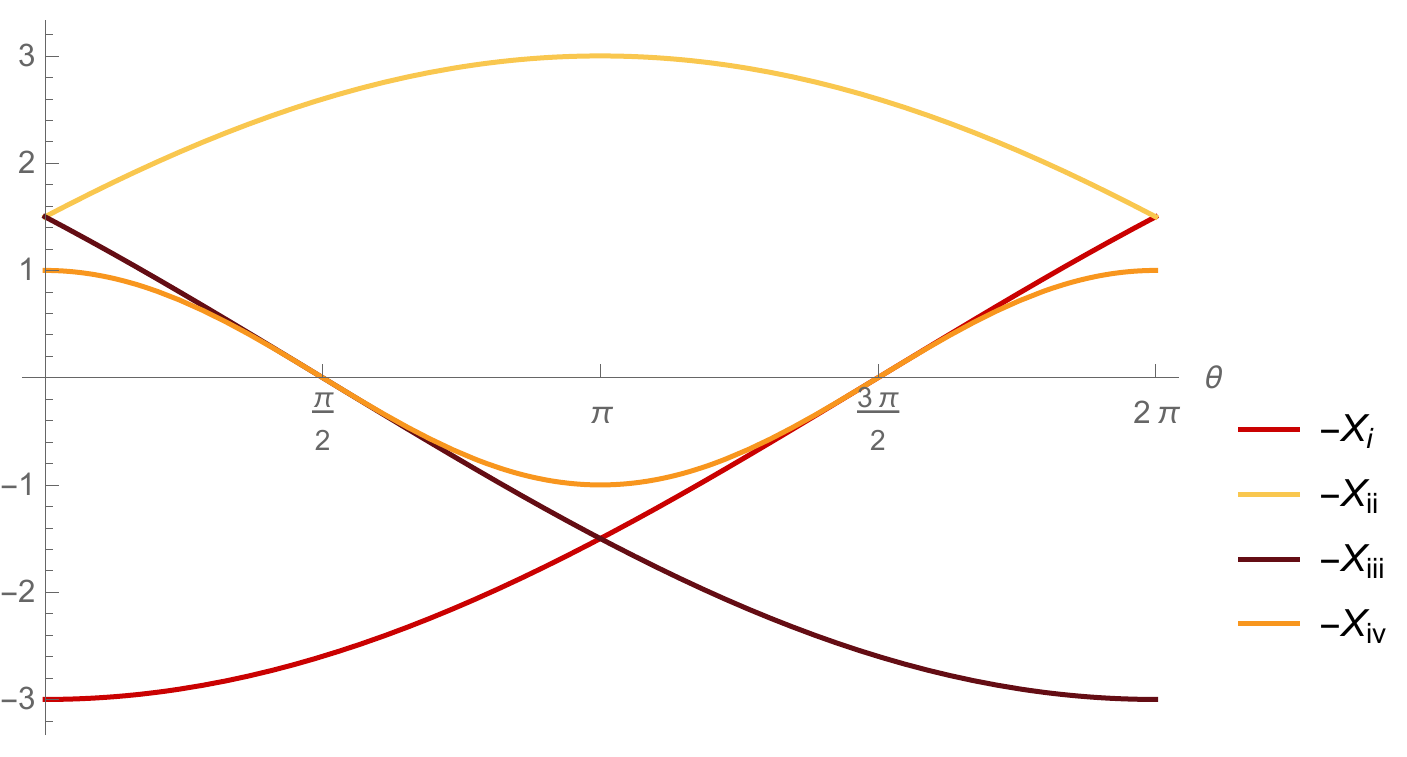} }}%
    \quad
    \subfloat[\centering $\theta$-dependence of the energy in the superfluid phase for $N_f=3$.]{{\includegraphics[scale=0.58]{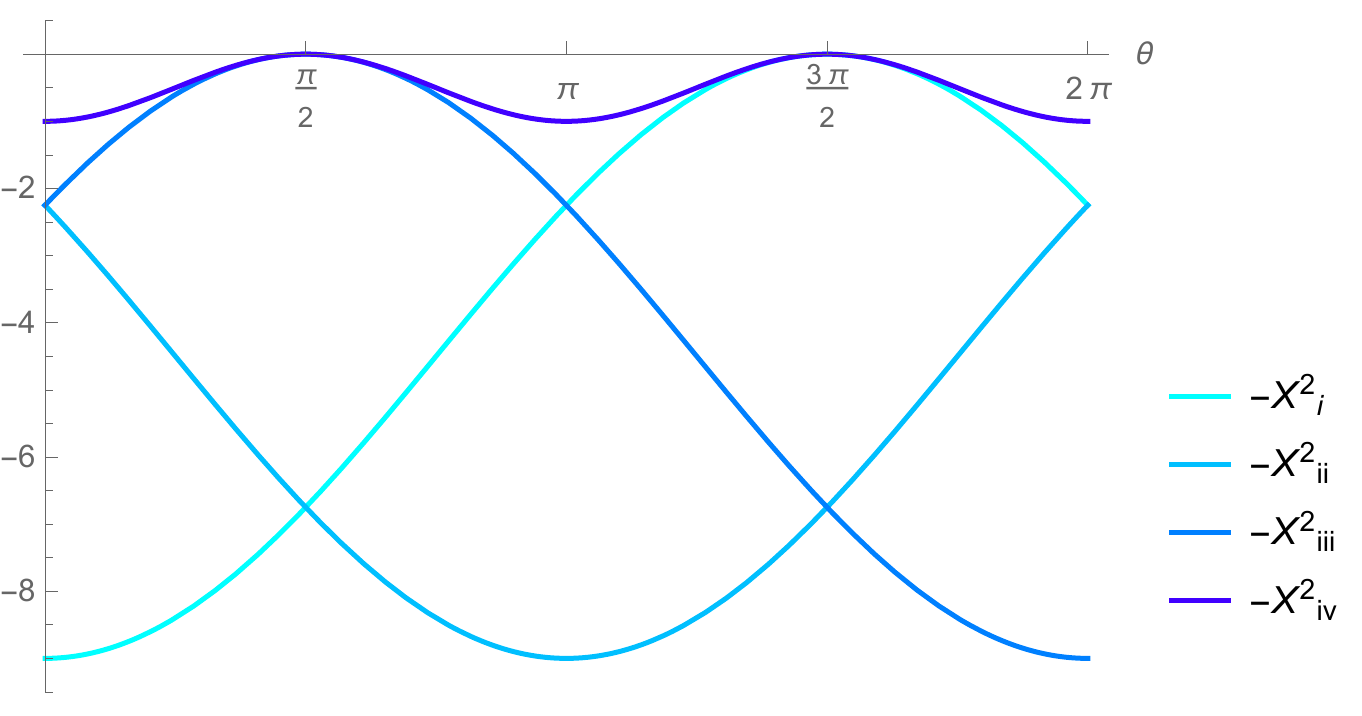} }}%
    \caption{$\theta$-dependence of the energy for $N_f=3$.}%
    \label{fig:ennodilN3}%
\end{figure}

\noindent
In Fig.\ref{fig:ennodilN3}.(a) we show the value of the variable $-X$ as a function of $\theta$. Recalling that in the normal phase, the energy is minimized when $X$ is maximized, we observe that the physical vacuum is ruled by the solutions $\mathbf{i.}$ and $\mathbf{iii.}$ which cross at $\theta = \pi$ where Dashen's phenomenon occurs \cite{Smilga:1998dh}. In the superfluid phase the story changes since now the minimum of the energy is achieved when $X^2$ is maximized. The situation is depicted in Fig. \ref{fig:ennodilN3}.(b): here the relevant solutions are  $\mathbf{i.}$,  $\mathbf{ii.}$,  $\mathbf{iii.}$. Solutions $\mathbf{i.}$ and $\mathbf{ii.}$ cross at $\theta=\frac{\pi}{2}$ whereas $\mathbf{ii.}$ and $\mathbf{iii.}$ cross at $\theta=\frac{3\pi}{2}$. Therefore, at $\theta=\pi$ we have a unique minimum and $CP$ is conserved. In other words, Dashen's phenomenon at $\theta=\pi$ is again absent in the superfluid region. However, two new non-analytic points occur, one at $\theta=\frac{\pi}{2}$ and the other at $\theta=\frac{3\pi}{2}$. These are indications of two novel first-order phase transitions because the derivative of the free energy (corresponding to the $CP$ order parameter $\expval{F\widetilde{F}}$) jumps at these two values of $\theta$. \\
 
\begin{figure}[h!]
    \centering
    \includegraphics[scale=0.7]{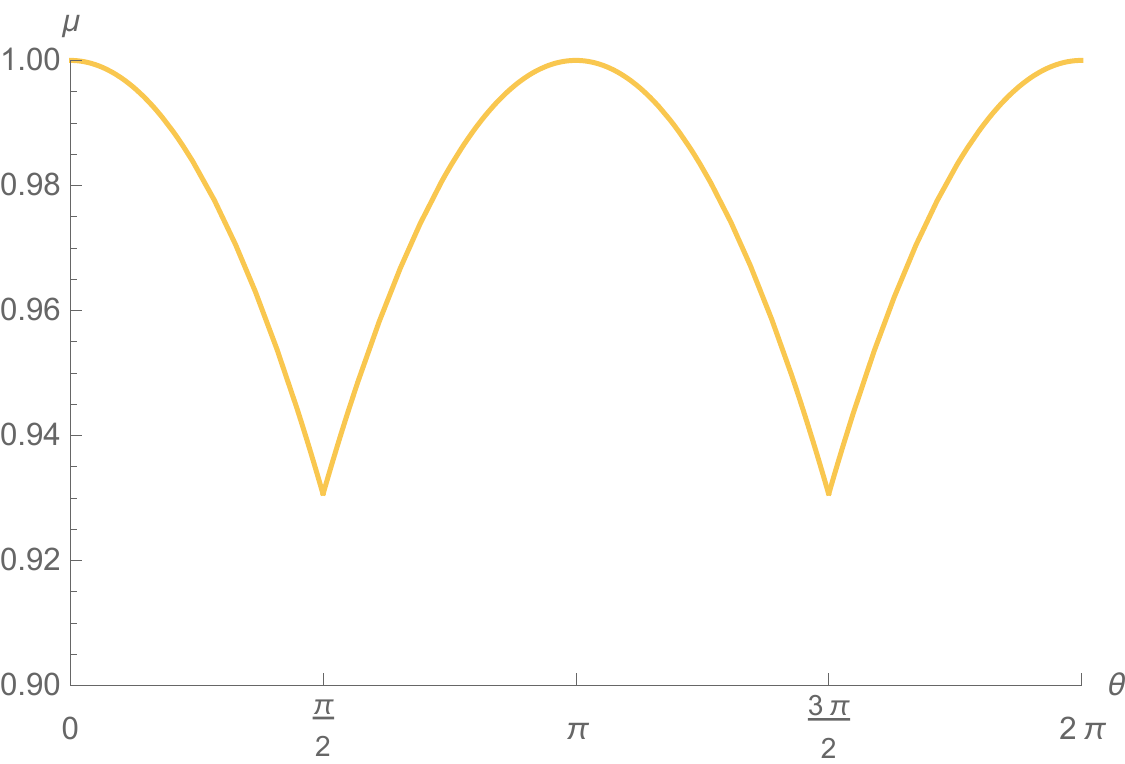}
     \caption{Normalized critical chemical potential ($m_\pi=\nu=1$) above which the superfluid phase is realized as a function of $\theta$ for $N_f=3$.}
    \label{fig:spN3}
\end{figure}

Having determined the $\theta$-vacuum in the two regions, we proceed by studying the transition between the normal and superfluid phases. Since the $U(\alpha_i)$ that minimizes the energy depends on $\theta$ and differs in the two phases, one should determine the critical $\mu$ by comparing the energies \eqref{en1senzadil} and \eqref{en2senzadil} in the given intervals of $\theta$.  We find that when the superfluid solution exists (i.e. $\cos \varphi \le 1$) it is always realized. In turn, the critical chemical potential is $\mu_c = m_\pi (\theta)$, which is displayed in Fig.\ref{fig:spN3}. Differently from the $N_f=2$ case, $\mu_c$ exhibits a mild dependence on $\theta$ and oscillates near the value $\mu_c = m_\pi$.

\begin{figure}[h!]
    \centering
    \subfloat[\centering The $CP$ order parameter $\expval{F \Tilde{F}}$ as a function of $\theta$ in the normal phase. We consider $m_\pi=\nu=1$.]{{\includegraphics[scale=0.7]{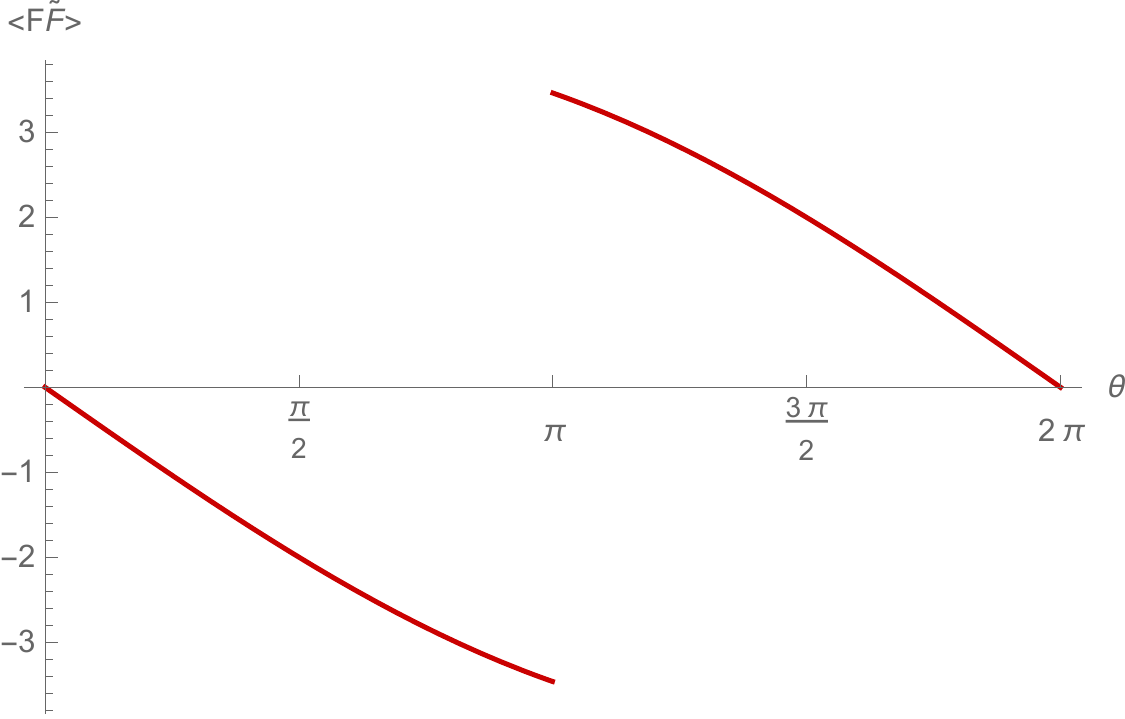} }}%
    \quad
    \subfloat[\centering  The $CP$ order parameter $\expval{F \Tilde{F}}$ as a function of $\theta$ in the superfluid phase for different values of the chemical potential: $\mu=5,10, 15$. We consider $m_\pi=\nu=1$. ]{{\includegraphics[scale=0.7]{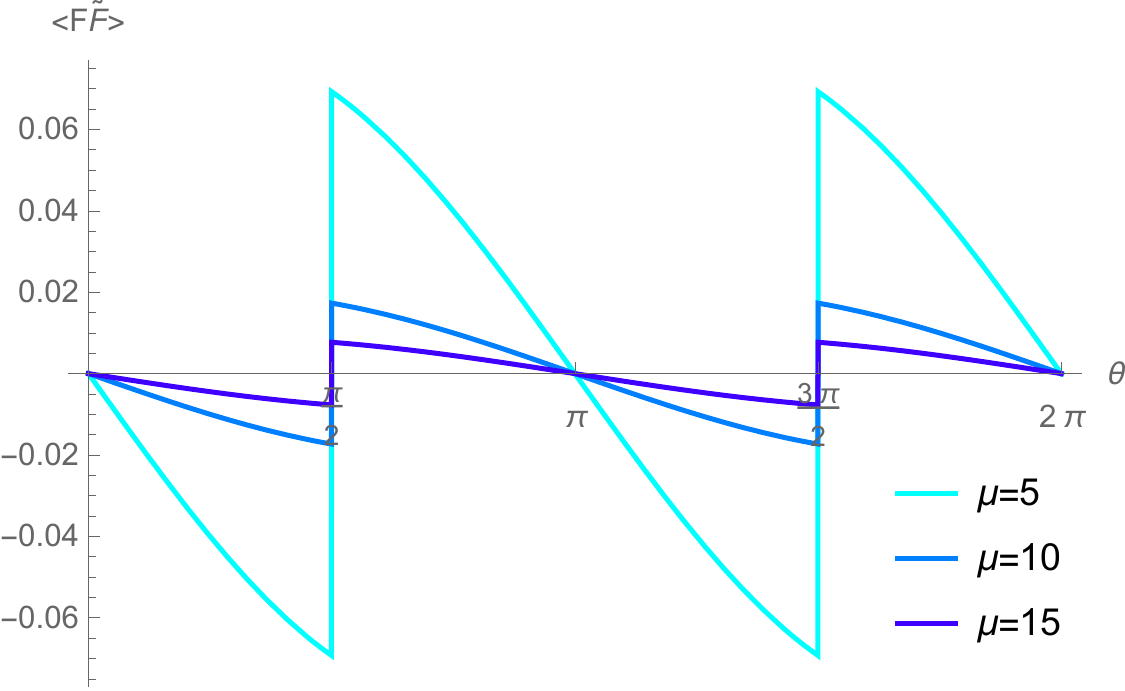} }}%
    \caption{$\theta$-dependence of the $CP$ order parameter for $N_f=3$.}%
    \label{fig:deXnorm}%
\end{figure}

The $CP$ order parameter is plotted for the normal and superfluid phases in Fig.\ref{fig:deXnorm} (a) and Fig. \ref{fig:deXnorm} (b), respectively. 
Finally, we show the value of $\bar{\theta}$ at $\theta=\pi$ for the ground state energy. The expression in the $\frac{m_{\pi}^{2}}{a}$ expansion is
\begin{align}
    \bar{\theta}&=\frac{\sqrt{3} m_{\pi}^2}{a}-\frac{m_{\pi}^4}{\sqrt{3} a^2}-\frac{m_{\pi}^6}{6 \sqrt{3} a^3}+\cO\left(\frac{m_\pi^8}{a^4}\right),\qquad &\text{normal phase}\\
    \bar{\theta}&=0, \qquad &\text{superfluid phase}\ .
\end{align}

\subsection{$\mathrm{N_f = 6}$}

For  the case of $N_f=6$ we, at first, assume that chiral symmetry breaking occurs away from a potential nearby (as function of flavors) conformal phase that will be discussed at length in \cite{PartII}. The full set of solutions is given by
\begin{align} 
\label{N6}
    \text{Solutions i-vi}&: \alpha_1 = \alpha_2, \dots =\alpha_6=\frac{\theta+2\pi k}{6},\quad &k=0,\dots,5  \nonumber \\ 
    \text{Solutions vii-ix}&: \alpha_1 = \alpha_2 = \dots =\alpha_5=\frac{\theta-\pi+2\pi k}{4}\,, \quad  \alpha_6=\pi-\alpha_1 \,, \quad &k=0,\dots,3  \nonumber \\ 
     \text{Solutions x-xii}&: \alpha_1 =  \alpha_2 =\dots= \alpha_4=\frac{\theta-2\pi+2\pi k}{2}, \quad \alpha_5 = \alpha_6=\pi-\alpha_1,\ \quad &k=0,1\ .
\end{align}

\begin{figure}[h!]
    \centering
    \subfloat[\centering $\theta$-dependence of the energy for $N_f=6$ both in the normal and superfluid phase.]{{\includegraphics[scale=0.8]{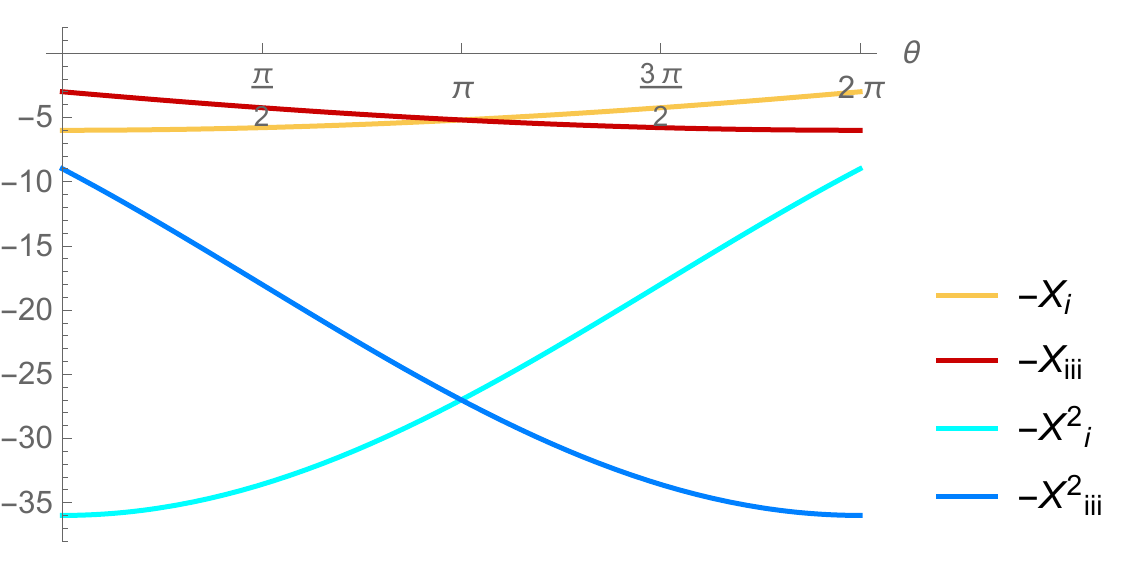} }}%
    \qquad
    \subfloat[\centering Normalized critical chemical potential ($m_\pi=\nu=1$) above which the superfluid phase is realized as a function of $\theta$.]{{\includegraphics[scale=0.6]{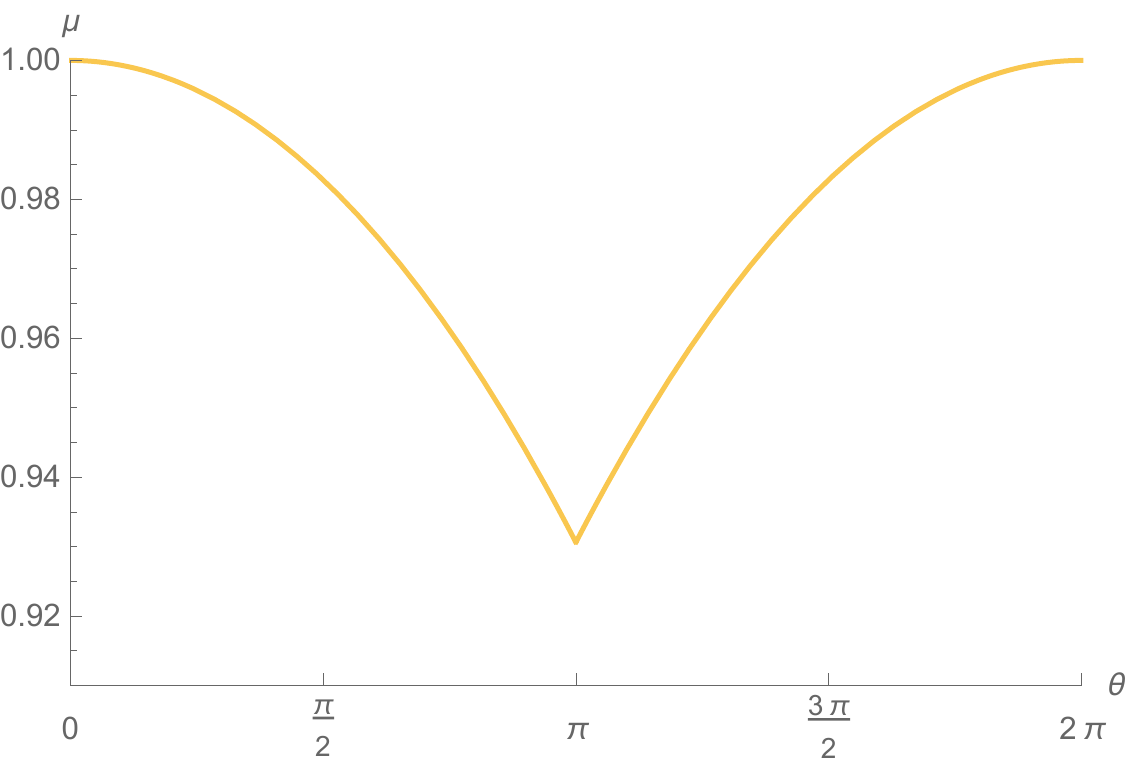} }}%
    \caption{$\theta$-dependence of the energy and of the critical chemical potential for $N_f=6$.}%
    \label{fig:enN6}%
\end{figure}

Here the relevant solutions for $U(\alpha_i)$ have $\alpha_1 =\alpha_2 = \dots = \alpha_6 = \alpha$. Moreover, all the solutions appear in pairs of degenerate energy in the superfluid phase as already observed in the $N_f=2$ case. This can be understood by noting that when $N_f$ is even, then from a given solution $\alpha$ of eq.\eqref{Solgen} it is possible to build another solution as $\alpha+\pi$ that will lead to the same $-X^2$.  As displayed in Fig. \ref{fig:enN6}.(a), in the normal phase the minimum of the energy is achieved for $\alpha = \frac\theta6$ (defined as solution i) and $\alpha = \frac{\theta+10 \pi}{6}$ (solution iii) with the two solutions crossing at $\theta= \pi$. This occurs also in the superfluid phase but now solution i and iii are degenerate for every value of $\theta$ with $\alpha= \frac{\theta}{6}+\pi$ and $\alpha = \frac{\theta+4\pi}{6}$, respectively. In Fig. \ref{fig:enN6}.(b) we show the phase diagram of the theory as a function of $\theta$ and $\mu$.  Differently from  $N_f =2$ we still observe a Dashen phenomenon at $\theta = \pi$ and therefore  $CP$  breaks spontaneously for this value. 

\begin{figure}[h!]
    \centering
    \subfloat[\centering The $CP$ order parameter $\expval{F \Tilde{F}}$ as a function of $\theta$ in the normal phase.  We consider $m_\pi=\nu=1$.]{{\includegraphics[scale=0.7]{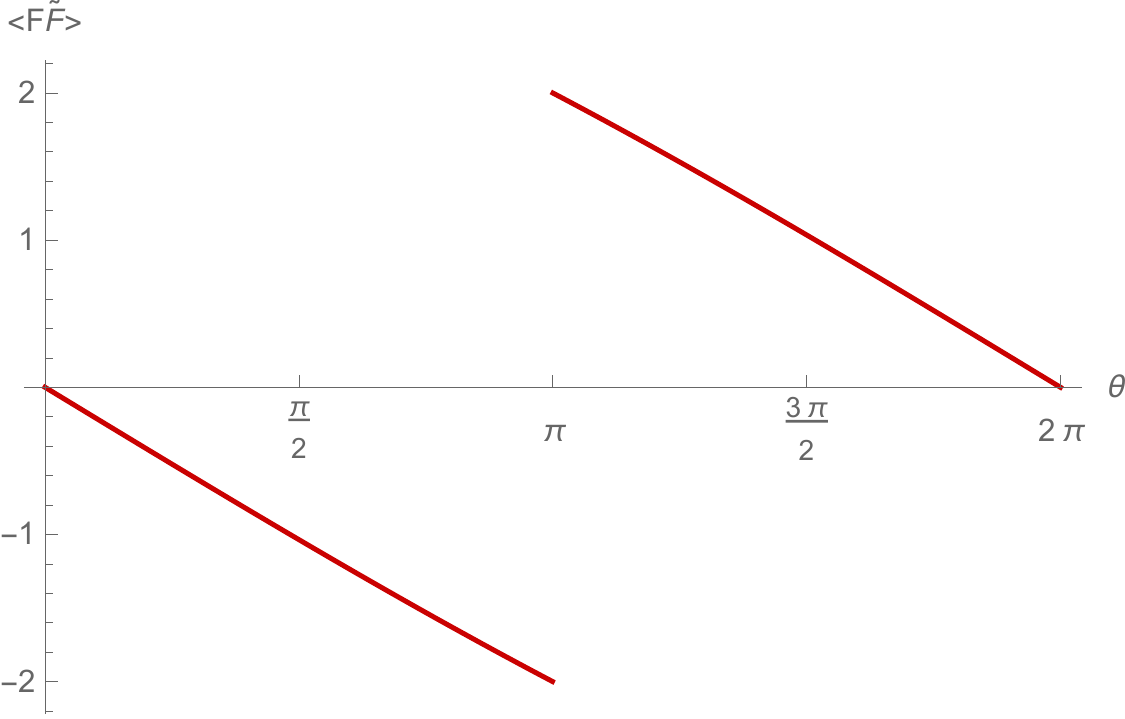} }}%
    \qquad
    \subfloat[\centering The $CP$ order parameter $\expval{F \Tilde{F}}$ as a function of $\theta$ in the superfluid phase for different values of the chemical potential: $\mu=5, 10, 15$.  We consider $m_\pi=\nu=1$.]{{\includegraphics[scale=0.7]{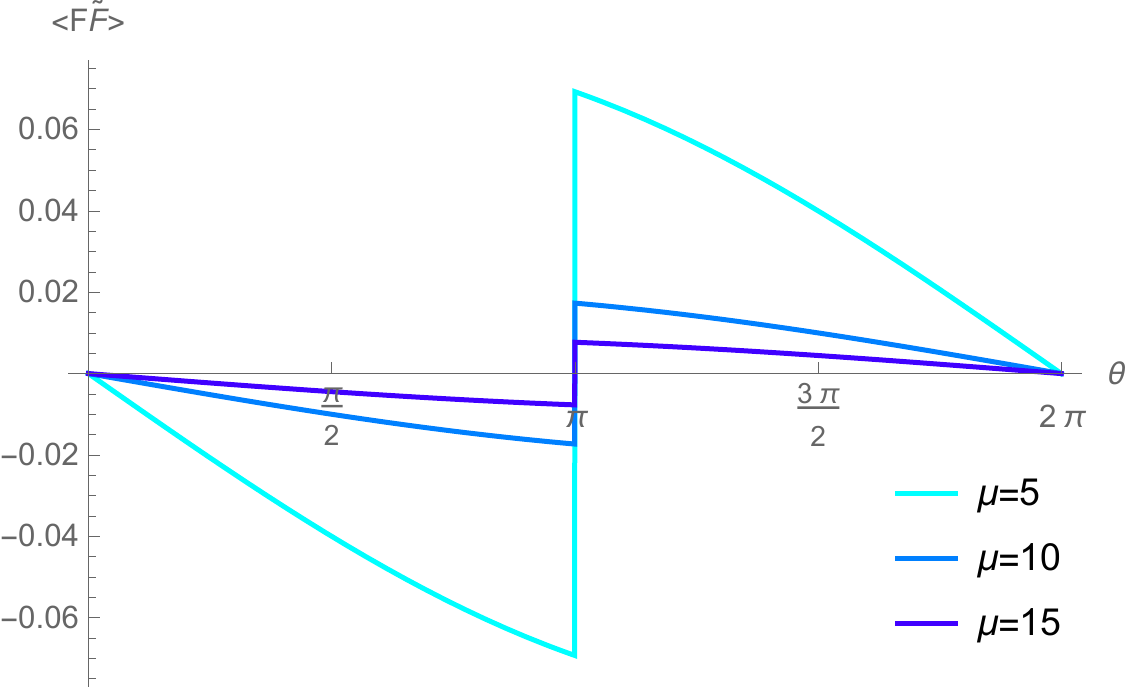} }}%
    \caption{$\theta$-dependence of the $CP$ order parameter for $N_f=6$.}%
    \label{fig:deXnormN6}%
\end{figure}

In the $N_f=6$ case we therefore have spontaneous breaking of $CP$ at $\theta=\pi$ in both phases as can be seen considering the $CP$ order parameter represented in Fig.\ref{fig:deXnormN6}.(a) and Fig. \ref{fig:deXnormN6}.(b).
\\
Finally, we report the value of $\bar{\theta}$ at $\theta=\pi$ expanded in $\frac{m_{\pi}^{2}}{a}$ 
\begin{align}
    \bar{\theta}&=\frac{m_{\pi}^2}{a}-\frac{m_{\pi}^4}{2 \sqrt{3} a^2}+\frac{5 m_{\pi}^6}{72 a^3}+\cO\left(\frac{m_\pi^8}{a^4}\right),\quad &\text{normal phase}\\
    \bar{\theta}&=-\frac{\sqrt{3} m_{\pi}^4}{2 a \mu ^2}+\frac{m_{\pi}^8}{4 \sqrt{3} a^2 \mu ^4}+\frac{m_{\pi}^{12}}{48 \sqrt{3} a^3 \mu ^6}+\cO\left(\frac{m_\pi^{16}}{a^4\mu^8}\right), \qquad &\text{superfluid phase}\ .
\end{align}

\subsection{$\mathrm{N_f=7}$}

\begin{figure}[h!]
    \centering
    \subfloat[\centering $\theta$-dependence of the energy for $N_f=6$ both in the normal and superfluid phase.]{{\includegraphics[scale=0.8]{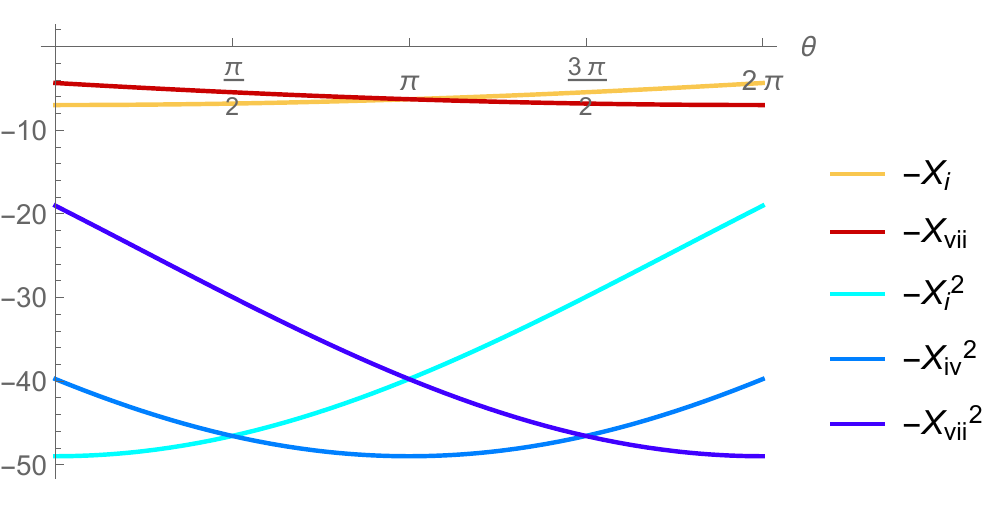} }}%
    \qquad
    \subfloat[\centering Normalized critical chemical potential ($m_\pi=\nu=1$) above which the superfluid phase is realized as a function of $\theta$.]{{\includegraphics[scale=0.6]{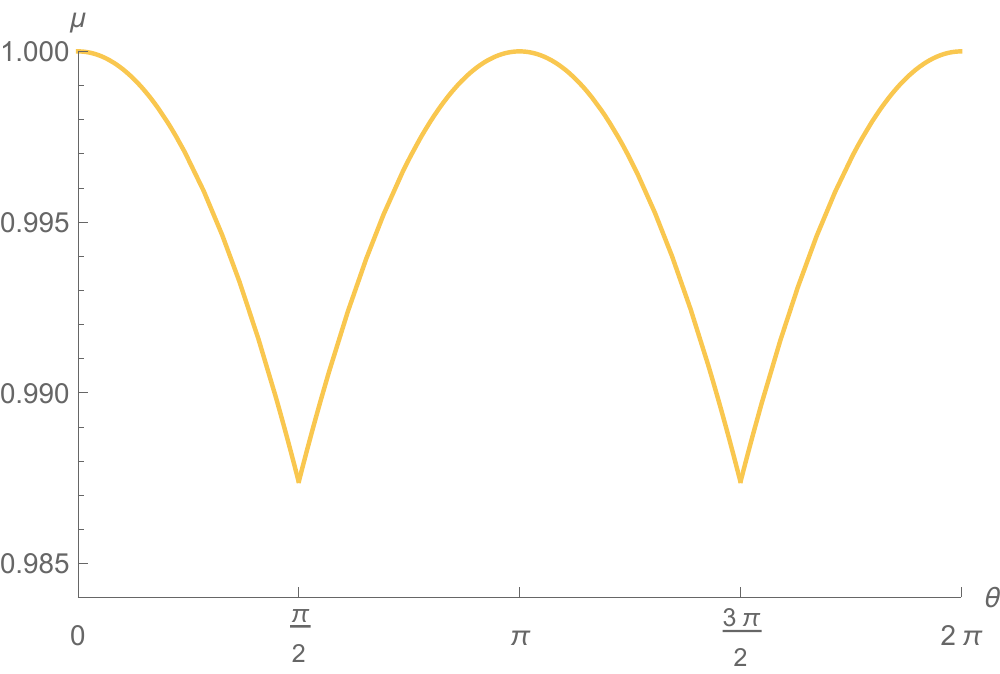}}}%
    \caption{$\theta$-dependence of the energy and of the critical chemical potential for $N_f=7$.}%
    \label{fig:min7}%
\end{figure}

 For the $N_f=7$ case we have thus $16$ solutions of eq.\eqref{Solgen} which are given by
\begin{align} 
\label{N7}
    \text{Solutions i-vii}&: \alpha_1 = \alpha_2, \dots =\alpha_7=\frac{\theta+2\pi k}{7},\quad &k=0,\dots,6  \nonumber \\ 
    \text{Solutions viii-xii}&: \alpha_1 = \alpha_2 = \dots =\alpha_5=\frac{\theta-\pi+2\pi k}{5}\,, \quad  \alpha_6= \alpha_7=\pi-\alpha_1 \,, \quad &k=0,\dots,4  \nonumber \\ 
     \text{Solutions xiii-xv}&: \alpha_1 =  \alpha_2 = \alpha_3=\frac{\theta-2\pi+2\pi k}{3}, \quad \alpha_4 = \alpha_5 = \alpha_6 =\alpha_7=\pi-\alpha_1,\ \quad &k=0,\dots,2 \nonumber \\
      \text{Solution xvi}&:
    \alpha_1=\theta-3\pi, \quad \alpha_2=\alpha_3=\dots=\alpha_7=\pi-\alpha_1  \ .
\end{align}
The solution that minimizes the energy and their corresponding $\theta$-dependence in the two phases is shown in Fig.\ref{fig:min7}. As in the previous cases, the solutions that minimize the energy have all equal angles, corresponding here to the first set of solutions in \eqref{N7}. In particular, in the normal phase the minimum of the energy is achieved for $\alpha = \frac{\theta}{7}$ and $\alpha = \frac{\theta+12 \pi}{7}$ with the two solutions crossing at $\theta= \pi$ (in Fig.\ref{fig:min7}.(a)). The $\theta$-dependence in the superfluid phase is analogous to the $N_f=3$ case. In fact, the relevant solutions are $\alpha= \frac{\theta}{7}$, $\alpha = \frac{\theta+6\pi}{7}$, and $\alpha=\frac{\theta+12 \pi}{7}$ with the first two crossing at $\theta=\frac{\pi}{2}$ while the second and the third at $\theta=\frac{3\pi}{2}$. As already described in the $N_f=3$ section, the energy is characterised by a unique minimum and $CP$ intact symmetry at $\theta=\pi$. 
\\

\begin{figure}[h!]
    \centering
    \subfloat[\centering The $CP$ order parameter $\expval{F \Tilde{F}}$ as a function of $\theta$ in the normal phase.  We consider $m_\pi=\nu=1$.]{{\includegraphics[scale=0.7]{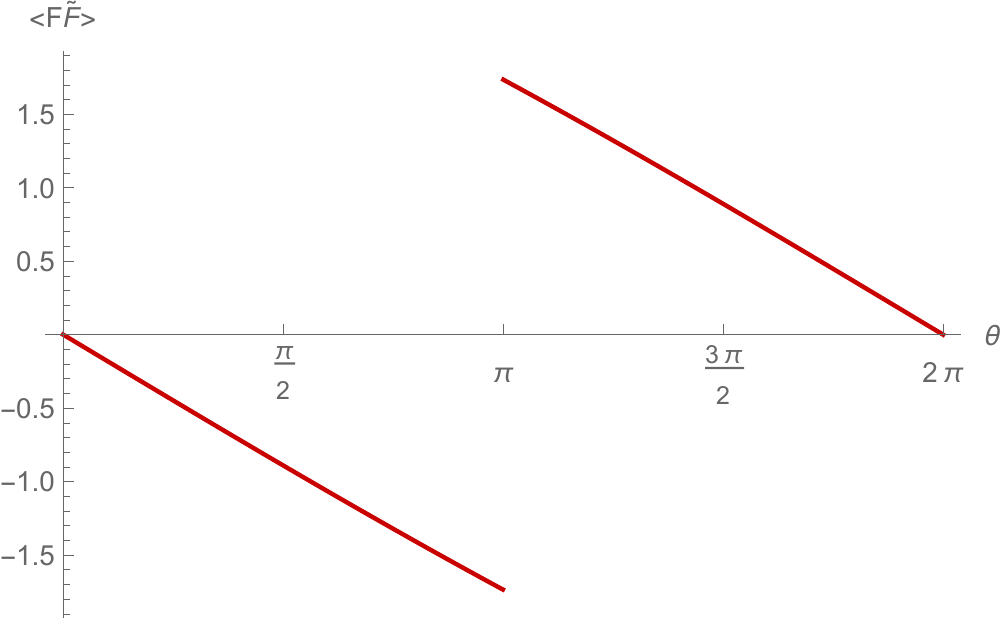} }}%
    \qquad
    \subfloat[\centering The $CP$ order parameter $\expval{F \Tilde{F}}$ as a function of $\theta$ in the superfluid phase for different values of the chemical potential: $\mu=5, 10, 15$.  We consider $m_\pi=\nu=1$.]{{\includegraphics[scale=0.7]{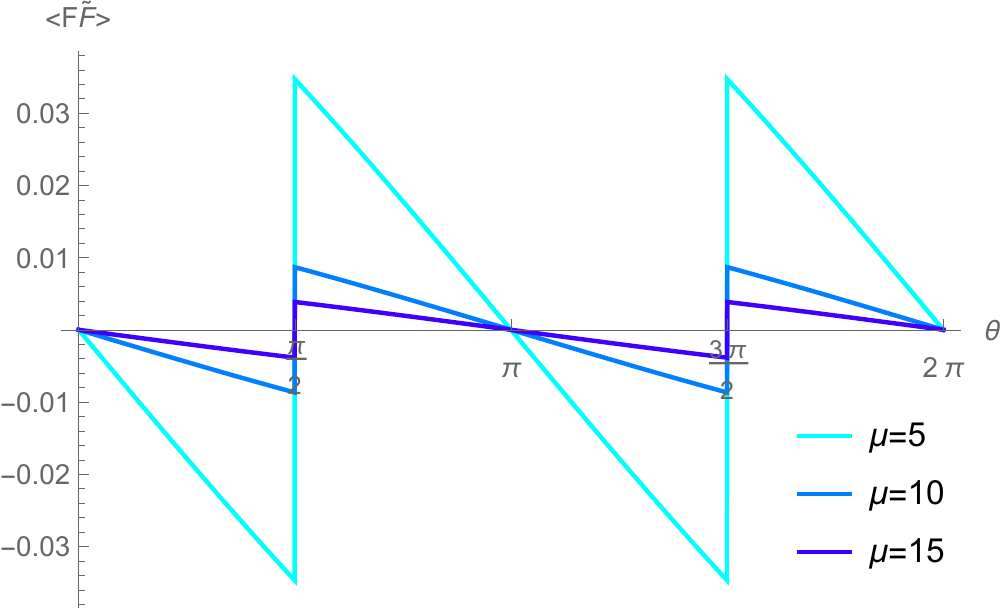} }}%
    \caption{$\theta$-dependence of the $CP$ order parameter for $N_f=7$.}%
    \label{fig:deXnormN7}%
\end{figure}

The critical chemical potential is displayed in Fig.\ref{fig:min7}.(b). It oscillates with period $2 \pi$ between the values $\mu_c = m_\pi$ and $\mu_c = m_\pi(\theta)$.
Fig. \ref{fig:min7} leads us to further investigate the physical meaning of the $\theta=\frac{\pi}{2}$ and $\theta=\frac{3\pi}{2}$ points. As a consequence, we analysed the $CP$ order parameter which we show in Fig.\ref{fig:deXnormN7} as a function of $\theta$. As can be seen from the figure, the aforementioned points are actually points of discontinuity of the $CP$ order parameter therefore it signals the occurrence of a first-order phase transition. 
\\
Finally, we provide the values of $\bar{\theta}$ at $\theta=\pi$  in the $\frac{m_{\pi}^{2}}{a}$ expansion 
\begin{align}
    \bar{\theta}&=\frac{2 m_\pi^2 \sin \frac{\pi }{7}}{a}-\frac{2 m_{\pi}^4 \cos \frac{3 \pi }{14}}{7 a^2}+\frac{2 m_{\pi}^6 \sin\frac{\pi }{7} \left(1+3 \sin \frac{3 \pi }{14}\right)}{49 a^3} + \cO\left( \frac{m_\pi^8}{a^4} \right),\qquad &\text{normal phase}\\
    \bar{\theta}&=0,\qquad &\text{superfluid phase}\ .
\end{align}

\subsection{$\mathrm{N_f = 8}$}
We proceed with the general expression for the solutions of \eqref{Solgen} for the $N_f=8$ case
\begin{align}
  \text{Solutions i-viii}&:  \alpha_1=\dots=\alpha_8=\frac{\theta+2\pi k}{8},\quad &k=0,\dots,7\\ 
  \text{Solutions ix-xiv}&: \alpha_1=\dots=\alpha_6=\frac{\theta-\pi+2\pi k}{6},  \quad\alpha_7=\alpha_8=\pi-\alpha_1,\quad &k=0,\dots,5\\
    \text{Solutions xv-xviii}&: \alpha_1=\dots=\alpha_4=\frac{\theta-2\pi+2\pi k}{4},  \quad \alpha_5=\dots=\alpha_8=\pi-\alpha_1,\quad &k=0,\dots,3\\
    \text{Solutions xix-xx}&: \alpha_1=\alpha_2=\frac{\theta-3\pi+2\pi k}{2}, \quad \alpha_3=\dots = \alpha_8=\pi-\alpha_1,\quad &k=0,1\ .
\end{align}

\begin{figure}[h!]
    \centering
    \subfloat[\centering $\theta$-dependence of the energy for $N_f=8$ both in the normal and superfluid phase.]{{\includegraphics[scale=0.8]{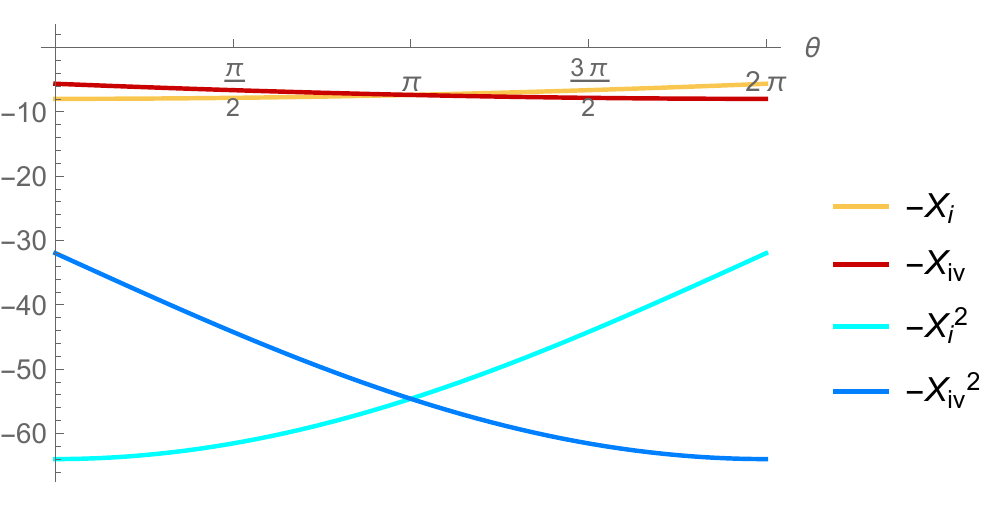} }}%
    \qquad
    \subfloat[\centering Normalized critical chemical potential ($m_\pi=\nu =1$) above which the superfluid phase is realized as a function of $\theta$.]{{\includegraphics[scale=0.6]{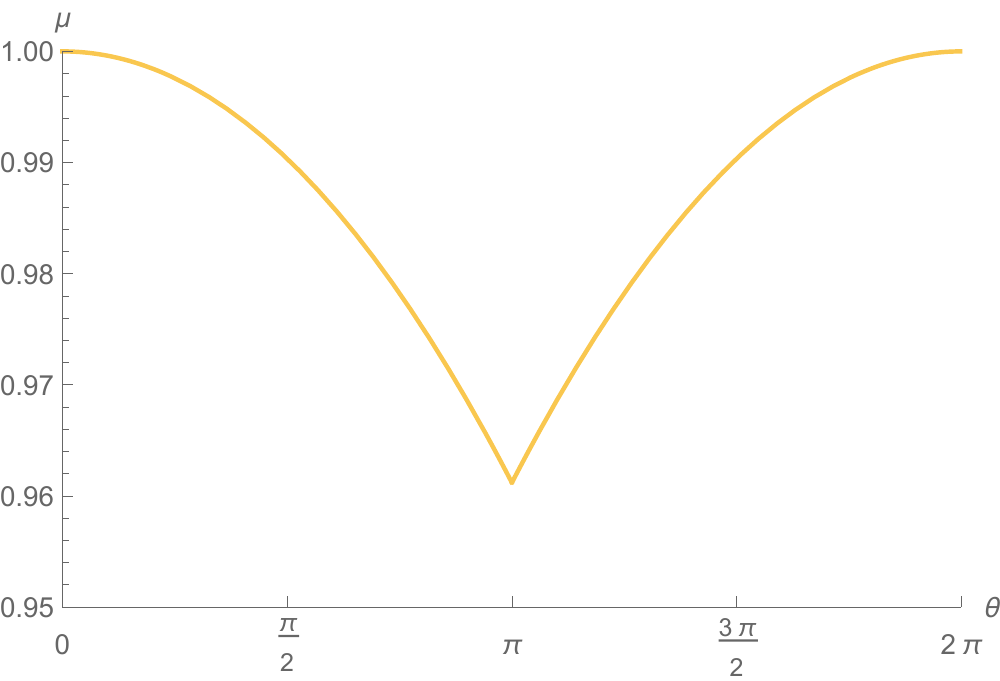} }}%
    \caption{$\theta$-dependence of the energy and of the critical chemical potential for $N_f=8$.}%
    \label{fig:min8}%
\end{figure}

\begin{figure}[h!]
    \centering
    \subfloat[\centering The $CP$ order parameter $\expval{F \Tilde{F}}$ as a function of $\theta$ in the normal phase.  We consider $m_\pi=\nu=1$.]{{\includegraphics[scale=0.7]{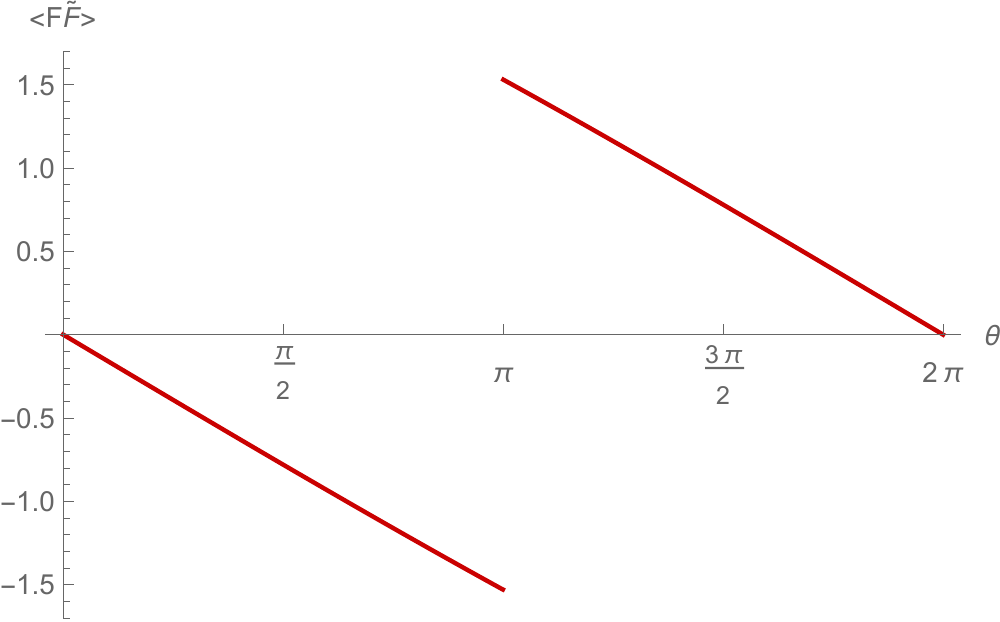} }}%
    \qquad
    \subfloat[\centering The $CP$ order parameter $\expval{F \Tilde{F}}$ as a function of $\theta$ in the superfluid phase for different values of the chemical potential: $\mu=5, 10, 15$.  We consider $m_\pi=\nu=1$.]{{\includegraphics[scale=0.7]{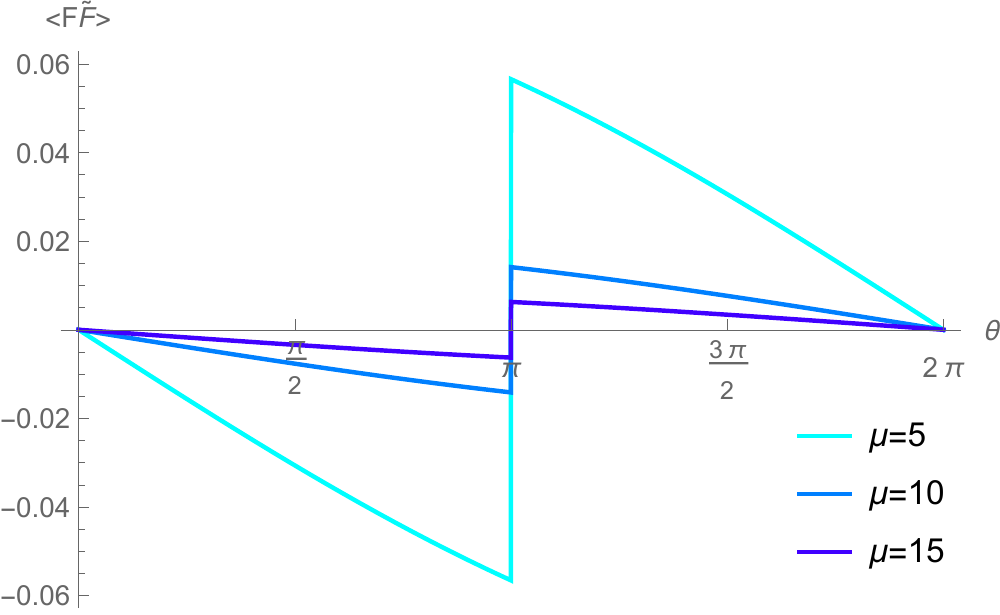} }}%
    \caption{$\theta$-dependence of the $CP$ order parameter for $N_f=8$.}%
    \label{fig:deXnormN8}%
\end{figure}

In line with our previous analyses, the $\theta$-dependence of the energy is minimised by the solutions with all equal angles. Fig. \ref{fig:min8} displays the behaviour of this minimum in the range $\theta\in\comm{0}{2 \pi}$ as well as the phase diagram of the theory. The $\theta$-dependence is analogous to $N_f=6$ case as it can be further deduced by studying  the $CP$ order parameter with its behaviour shown in Fig.\ref{fig:deXnormN8}
\\
The value of $\bar{\theta}$ at $\theta=\pi$  in an $\frac{m_{\pi}^{2}}{a}$ expansion reads
\begin{align}
    \bar{\theta}&=\frac{2 m_{\pi}^2 \sin \frac{\pi }{8}}{a}-\frac{m_{\pi}^4}{4 \sqrt{2} a^2}+ \frac{\left(1+\frac{3}{\sqrt{2}}\right) m_{\pi}^6 \sin \frac{\pi }{8}}{32 a^3} + \cO\left( \frac{m_\pi^8}{a^4} \right),\qquad &\text{normal phase}\\
    \bar{\theta}&=\frac{m_{\pi}^4}{\sqrt{2} a \mu ^2}-\frac{m_{\pi}^8}{8 a^2 \mu ^4}+\frac{m_{\pi}^{12}}{64 \sqrt{2} a^3 \mu ^6}+\cO\left(\frac{m_\pi^{16}}{a^4\mu^8}\right),\qquad &\text{superfluid phase}\ .
\end{align}

\subsection{Considerations for general $\mathrm{N_f}$}

Having analyzed in detail the physics for different values of $N_f$ away from the conformal window, we take a step back and consider the emerging structure hinting at structural differences between the even/odd $N_f$ cases in the superfluid phase.  We start by noting that solutions of the EOM are generally not periodic of $2 \pi$ for $\theta$. In fact, the periodicity condition can be satisfied only if at least two solutions cross in the interval $\theta \in [0, 2 \pi]$.  Taking $U=e^{-i \alpha} \identity_{2 N_f}$, one can ask when two different solutions of the equation of motion can have the same energy. This corresponds to  requiring 
\begin{align}
\cos\left(\frac{\theta+2 \pi k_1}{N_f}\right) &= \cos\left(\frac{\theta+2 \pi k_2}{N_f}\right)\ ,   & \text{normal phase} \\
\cos^2\left(\frac{\theta+2 \pi k_1}{N_f}\right) &= \cos^2\left(\frac{\theta+2 \pi k_2}{N_f}\right)\ ,  & \text{superfluid phase}\ .
\label{Con}
\end{align}
 Both conditions are satisfied when $k_1=-\frac{\theta}{\pi}-k_2+N_f$. Since $k_1$, $k_2$ are integers, it follows that the energy can only be non-trivially equal for $\theta=\pi$. Therefore to find the two different solutions, in the normal phase, covering the full  $[0, 2 \pi]$ interval for $\theta$ it is sufficient to consider the case $k_1=0$ that for $[0, \pi]$ interval corresponds to the ground state energy, furthermore at $\theta = 
\pi$  it forces the second solution to be $k_2 = N_f-1$. This result is in agreement with the findings for the specific cases above for the normal phase.  

In the superfluid phase, we have two additional solutions to the second condition  \eqref{Con}:
\begin{itemize}
    \item When $N_f$ is even we have the solution $k_1 = k_2 + \frac{N_f}{2}$ which does not depend on $\theta$. This corresponds to the observation that the solutions for even $N_f$ organize themselves in pairs ($\alpha$ and $\alpha+\pi$) with the same energy for every $\theta$. Additionally, of course, we still have the normal phase solution  $k_1 = 0$ and $k_2 = N_f-1$ together with their superfluid degenerate partners. These two classes of solutions cross at $\theta=\pi$ except for the special  $N_f=2$ case for which the crossing disappears. Note that the pairs of completely degenerate solutions remain such to all orders in $\frac{m_\pi^2}{a}$. In fact, given the EOM for a certain $\alpha$
    \be
    \frac{m_\pi^4}{a \mu ^2}\sin (2 \alpha ) = \theta - N_f \alpha  \,,
    \ee
    we have the same EOM for $\alpha + \pi$, upon shifting the $\theta$-angle as $\theta \to \theta + N_f \pi$. Being $N_f$ even, this corresponds to a shift by an integer multiple of $2 \pi$. 
    
    \item When $N_f$ is odd we have the solution $k_1 = -k_2 +\frac{N_f}{2} - \frac{\theta}{\pi}$ which can be realized  for $\theta=\frac{\pi}{2}$ and $\theta=\frac{3\pi}{2}$. Since the solution that minimizes the energy near $\theta=0$ is $\alpha = \frac{\theta}{N_f}$, then it will cross at $\theta=\frac{\pi}{2}$ with the solution  for $\theta >\frac{\pi}{2} $ that becomes $\alpha =  \frac{\theta +2\pi k_2 }{N_f}=\frac{\theta -\pi }{N_f}+\pi$. The latter solution represents the minimum till $\theta=\frac{3\pi}{2}$ where its energy crosses with $\alpha=\frac{\theta-2 \pi}{N_f}$. As a consequence, the vacuum is always non degenerate at $\theta=\pi$ where $CP$ is conserved. In particular, for $\theta=\pi$, $\bar \theta$ vanishes to all orders in $\frac{ m_\pi^4}{a \mu ^2}$ since the equation of motion \eqref{vensup} admits the minimum energy solution $\alpha = \theta = \pi$. On the other hand, we have two novel first-order phase transitions at $\theta=\frac{\pi}{2}\,, \frac{3\pi}{2}$ due to a jump of the physical vacuum between the two minima and characterized by a discontinuous $CP$ order parameter.

\end{itemize}

Finally, we estimate the tension of the domain wall between the two degenerate vacua at $\theta = \pi$ for even $N_f$ in the superfluid phase. To this end, we consider the Lagrangian \eqref{lagtheta} and look for solutions $\alpha_i =\alpha_i(x)$ that interpolate between the two degenerate ground states\eqref{2fold} at $\theta= \pi$ with $x$ the coordinate orthogonal to the domain wall. By considering the ansatz \eqref{anscomp}, we obtain the tension of the wall as
\be
T = 2 \nu^2 \int_{-\infty}^{\infty} \ d x \left[ \sum _{i=1}^{N_f} \alpha _i'(x){}^2-\frac{m_{\pi }^4}{\mu ^2 N_f} \left(\sum _{i=1}^{N_f} \cos \left(\frac{\pi }{N_f}+\alpha _i(x)\right)\right)^2\right]
\ee
where we have shifted $\alpha_i$ as $\alpha_i \to \alpha_i + \frac{\pi}{N_f}$ such that the boundary conditions are $\alpha(-\infty) = 0$ and $\alpha(\infty) = \pi$. As discussed in \cite{Gaiotto:2017tne}, the boundary condition fixes $\alpha_i = \alpha$ for $i=1, \dots, N_f-1$ and $\alpha_{N_f} = -(N_f-1) \alpha$, leading to
\be
\small
T = 2 \nu^2 \int_{-\infty}^{\infty} \ d x \left[\left(N_f-1\right) N_f \alpha'(x)^2-\frac{m_\pi^4}{\mu ^2 N_f} \left(\left(N_f-1\right) \cos \left(\alpha(x) +\frac{\pi }{N_f}\right)+\cos \left(\frac{\pi }{N_f}- \left(N_f-1\right)\alpha(x) \right)\right)^2\right]
\ee
By introducing dimensionless coordinates as  $\bar{x} =x \frac{ m_\pi^2}{\mu}$, we obtain that regardless of the exact form of the wall's profile, its tension scales as
\be
T \approx \frac{\nu^2 m_\pi^2}{\mu}
\ee
and shows a suppression by a factor or $\frac{m_\pi}{\mu}$ compared to the normal phase result $T \approx \nu^2 m_\pi$ \cite{Gaiotto:2017tne}.

\subsection{Ground state energy with the axion}

The effective action for two-color QCD at finite baryon number in the presence of the Peccei-Quinn axion field $N$ is
\begin{equation}
     L_{\text{eff}}=-a \nu ^2 \left (\theta -\sum_{i=1}^{N_f} \alpha_i-\delta \right)^2-\Lambda ^4+4 \nu ^2 m_\pi^2 X \cos \varphi +2 \mu ^2 \nu ^2 N_f \sin ^2\varphi  \ , \qquad  
\langle N \rangle = e^{-i\delta/a_{PQ}} 
\ ,
\end{equation}
whose equation of motion $\theta -\sum_{i=1}^{N_f} \alpha_i-\delta=0$ is solved for $\delta=\theta$ and $\alpha_i=0$. This solution minimizes the energy and leads to $X=N_f$. Hence, the superfluid ground state energy in presence of the axion corresponds to the known result at $\theta=0$ \cite{kogut2000qcd}
\begin{equation}
    E=  -\frac{2 \nu^2
    N_f }{\mu
   ^2} \left(m_{\pi }^4+\mu ^4\right)\ .
\end{equation}

\section{Symmetry-breaking pattern and spectrum}
\label{SBP}
In the section above we concentrated on determining the vacuum structure of the theory. We now move to establish the associated symmetry-breaking pattern starting with the theory without an axion. The pattern will allow us to disentangle the light degrees of freedom of the theory. We will keep the analysis as general as possible so that it is applicable also for the Part II of our work related to near-conformal field theories in the large-charge expansion \cite{PartII}. 
\subsection{Symmetry-breaking pattern without the axion}
\label{SBPWA}

 The dynamics of the theory at hand depends on three parameters: the mass of the quarks $m$, the chemical potential $\mu$, and the scale of the axial anomaly $a$. The latter corresponds to the energy at which the pseudo-Goldstone (the $S$-particle) emerging from the spontaneous symmetry breaking of the axial symmetry acquires an anomalous mass. We are interested in the low energy theory where $S$ is kept into the spectrum. We, therefore, consider the following hierarchy \footnote{This hierarchy of scales is typical when modelling neutron stars in $2$-color QCD at finite isospin chemical potential \cite{Gandolfi:2019zpj}.}
\begin{equation} \label{scales}
    m \ll \sqrt{a} \leq \mu  \ll 4\pi \nu\ ,
\end{equation}
where the last inequality implies the validity of the chiral EFT. For $m=\mu=0$ and in absence of the singlet $S$, the infrared spectrum consists of massless Goldstone bosons of the spontaneously broken chiral symmetry as summarized below.
\\ 

\tikzset{every picture/.style={line width=0.75pt}} 

\begin{tikzpicture}[x=0.75pt,y=0.75pt,yscale=-1,xscale=1]

\draw [color={rgb, 255:red, 100; green, 13; blue, 20 }  ,draw opacity=1 ]   (428,136) .. controls (443.36,132.93) and (456.88,114.56) .. (460.59,99.82) ;
\draw [shift={(461,98)}, rotate = 101.31] [color={rgb, 255:red, 100; green, 13; blue, 20 }  ,draw opacity=1 ][line width=0.75]    (6.56,-1.97) .. controls (4.17,-0.84) and (1.99,-0.18) .. (0,0) .. controls (1.99,0.18) and (4.17,0.84) .. (6.56,1.97)   ;

\draw (58.6,26.8) node [anchor=north west][inner sep=0.75pt]  [font=\normalsize]  {$ \begin{array}{l}
m\ =\ 0\\
\mu =\ 0\\
\sqrt{a} \gg \ \nu 
\end{array}$};
\draw (285.4,129.2) node [anchor=north west][inner sep=0.75pt]  [font=\normalsize]  {$SU( 2N_{f})\overset{\textcolor[rgb]{0.39,0.05,0.08}{2N}\textcolor[rgb]{0.39,0.05,0.08}{_{f}^{2}}\textcolor[rgb]{0.39,0.05,0.08}{-N}\textcolor[rgb]{0.39,0.05,0.08}{_{f}}\textcolor[rgb]{0.39,0.05,0.08}{-1}}{\rightsquigarrow } Sp( 2N_{f}) \ $};
\draw (428.8,31.4) node [anchor=north west][inner sep=0.75pt]  [font=\normalsize]  {$\textcolor[rgb]{0.39,0.05,0.08}{2N}\textcolor[rgb]{0.39,0.05,0.08}{_{f}^{2}}\textcolor[rgb]{0.39,0.05,0.08}{-N}\textcolor[rgb]{0.39,0.05,0.08}{_{f}}\textcolor[rgb]{0.39,0.05,0.08}{-1}$};
\draw (349,57.2) node [anchor=north west][inner sep=0.75pt]  [font=\normalsize] [align=left] {Goldstones transforming under the\\ antisymm. representation of $\displaystyle Sp( 2N_{f})$};

\end{tikzpicture}\\\newline 
 When setting to zero the anomaly, and in the absence of the quark mass, the $S$-field becomes the Goldstone boson of the $U(1)_A$. 

\bigskip
We now provide the full spectrum of light particles  in the following limits

\begin{enumerate}
     \item \label{case2} \textbf{$\mathbf{m \neq 0, \mu =0}$ and $\mathbf{\sqrt{a} \ll  4\pi\nu}$} \\
Working in steps,  one first considers the would-be pattern of spontaneous symmetry breaking according to which $SU(2N_f)\times U(1)_A$ breaks spontaneously to $Sp(2N_f)$. Then the associated $2N_f^2 - N_f$ Goldstone bosons acquire masses via the non-zero quark mass and the anomaly.   
         \item \label{case3} \textbf{$\mathbf{m=0, \mu\not=0}$ and $\mathbf{\sqrt{a} \ll 4\pi \nu}$}\\
    The introduction of the chemical potential breaks explicitly the final $Sp(2 N_f)$ to $SU(N_f)_L\times SU(N_f)_R\times U(1)_B$ while the anomaly breaks $U(1)_A$. The relativistic Bose-Einstein condensation leads to the further spontaneous symmetry breaking $SU(N_f)_L \times SU(N_f)_R \times U(1)_B \rightsquigarrow Sp(N_f)_L \times Sp(N_f)_R $. Therefore, the final  spectrum is composed of $N^2_f-N_f-1$ massless Goldstone bosons the massive $S$ state due to the anomaly, and $N^2_f$ modes with mass proportional to the chemical potential. The latter belong to the $(N_f,N_f)$ irreducible representation of $Sp(N_f)_L \times Sp(N_f)_R $ whereas the true Goldstones arrange themselves into three different irreducible representations: a singlet $(1,1)$, the $\left(\frac{N_f(N_f-1)}{2}-1,1\right)$ representation and the $\left(1,\frac{N_f(N_f-1)}{2}-1\right)$. The sum of the degrees of freedom adds to $2N^2_f-N_f$.

   \item \label{case5} \textbf{$\mathbf{m\not=0, \mu\not=0}$ and $\mathbf{\sqrt{a} \ll 4\pi
   \nu}$}\\
 The situation here is involved due to the competition between the quark mass term and the baryon chemical potential.  The chemical potential breaks $Sp(2 N_f)$ down to $SU(N_f)_L \times SU(N_f)_R \times U(1)_B$, which, in turn, is explicitly broken by the mass term to its vectorial subgroup $SU(N_f)_V \times U(1)_B$. Finally, the superfluid vacuum breaks the latter symmetry group to $Sp(N_f)_V$ yielding to $\frac{N_f^2-N_f}{2}$ massless Goldstone bosons as summarized below

\begin{center}

\tikzset{every picture/.style={line width=0.75pt}} 
\begin{tikzpicture}[x=0.75pt,y=0.75pt,yscale=-1,xscale=1]

\draw [color={rgb, 255:red, 202; green, 1; blue, 1 }  ,draw opacity=1 ]   (372.28,90.4) -- (372.28,114.91) ;
\draw [shift={(372.28,116.91)}, rotate = 270] [color={rgb, 255:red, 202; green, 1; blue, 1 }  ,draw opacity=1 ][line width=0.75]    (6.56,-1.97) .. controls (4.17,-0.84) and (1.99,-0.18) .. (0,0) .. controls (1.99,0.18) and (4.17,0.84) .. (6.56,1.97)   ;
\draw [color={rgb, 255:red, 248; green, 150; blue, 30 }  ,draw opacity=1 ]   (372.27,159.2) .. controls (373.99,160.81) and (374.04,162.48) .. (372.43,164.2) .. controls (370.82,165.92) and (370.87,167.58) .. (372.59,169.19) .. controls (374.31,170.8) and (374.37,172.47) .. (372.76,174.19) .. controls (371.15,175.91) and (371.2,177.58) .. (372.92,179.19) .. controls (374.64,180.8) and (374.7,182.47) .. (373.09,184.19) .. controls (371.48,185.91) and (371.53,187.57) .. (373.25,189.18) .. controls (374.97,190.79) and (375.02,192.46) .. (373.41,194.18) -- (373.54,198.01) -- (373.8,206) ;
\draw [shift={(373.87,208)}, rotate = 268.12] [color={rgb, 255:red, 248; green, 150; blue, 30 }  ,draw opacity=1 ][line width=0.75]    (7.65,-3.43) .. controls (4.86,-1.61) and (2.31,-0.47) .. (0,0) .. controls (2.31,0.47) and (4.86,1.61) .. (7.65,3.43)   ;
\draw [color={rgb, 255:red, 202; green, 1; blue, 1 }  ,draw opacity=1 ]   (248.27,84.27) -- (348.53,115.05) ;
\draw [shift={(350.44,115.64)}, rotate = 197.07] [color={rgb, 255:red, 202; green, 1; blue, 1 }  ,draw opacity=1 ][line width=0.75]    (6.56,-1.97) .. controls (4.17,-0.84) and (1.99,-0.18) .. (0,0) .. controls (1.99,0.18) and (4.17,0.84) .. (6.56,1.97)   ;

\draw (205.2,46.73) node [anchor=north west][inner sep=0.75pt]  [font=\normalsize]  {$SU( 2N_{f}) \times U( 1)_{A}\overset{\textcolor[rgb]{0.39,0.05,0.08}{2N}\textcolor[rgb]{0.39,0.05,0.08}{_{f}^{2}}\textcolor[rgb]{0.39,0.05,0.08}{-N}\textcolor[rgb]{0.39,0.05,0.08}{_{f}}}{\rightsquigarrow } Sp( 2N_{f}) \ $};
\draw (301.4,128.93) node [anchor=north west][inner sep=0.75pt]  [font=\normalsize]  {$\textcolor[rgb]{0.79,0,0}{SU}\textcolor[rgb]{0.79,0,0}{(}\textcolor[rgb]{0.79,0,0}{N}\textcolor[rgb]{0.79,0,0}{_{f}}\textcolor[rgb]{0.79,0,0}{)}\textcolor[rgb]{0.79,0,0}{_{V}}\textcolor[rgb]{0.79,0,0}{\times U}\textcolor[rgb]{0.79,0,0}{(}\textcolor[rgb]{0.79,0,0}{1}\textcolor[rgb]{0.79,0,0}{)}\textcolor[rgb]{0.79,0,0}{_{B}} \ $};
\draw (380.28,90.75) node [anchor=north west][inner sep=0.75pt]    {$\textcolor[rgb]{0.79,0,0}{\mu }$};
\draw (336.6,216.4) node [anchor=north west][inner sep=0.75pt]  [font=\normalsize]  {$Sp( N_{f})_{V} \ $};
\draw (385.6,164.73) node [anchor=north west][inner sep=0.75pt]  [font=\footnotesize]  {$\textcolor[rgb]{0.97,0.59,0.12}{\frac{N_{f}^{2} -N_{f}}{2}}$};
\draw (240.2,163.33) node [anchor=north west][inner sep=0.75pt]  [font=\small] [align=left] {superfluid phase};
\draw (273.2,98.1) node [anchor=north west][inner sep=0.75pt]  [font=\footnotesize,rotate=-16.73]  {$\textcolor[rgb]{0.79,0,0}{a,\ m}$};

\end{tikzpicture}
\end{center}

\end{enumerate}

The case of our interest is the last one, where we have $\frac{N^2_f-N_f}{2}$ Goldstones transforming according to the antisymmetric representation of $Sp(N_f)_V$ plus a singlet. Additionally, there's the $\eta^\prime$ like state $S$  with mass of order $\sqrt{a}$. 

The naive counting above of the number of Goldstone bosons holds correct here since there are no Goldstone bosons of type II \cite{Nielsen:1975hm} in this case. The reason is that the dimension of the representation of the Goldstone bosons is identical to the number of broken generators (see discussion in \cite{Orlando:2020yii}).

\subsection{Symmetry-breaking pattern with the axion}
\label{sbpaxion}

In this section, we analyse the symmetry-breaking pattern for $m \neq 0$, $\mu \neq 0$, $a \ll 
4\pi \nu$ but including  the presence of the additional $U(1)_{PQ}$ symmetry discussed at the end  of Sec.\ref{qcdsymmetries}. The latter is classically exact but quantum mechanically anomalous and spontaneously broken, leading to the existence of a new pseudo-Goldstone boson (the axion) with the mass of the order of the scale of the anomaly. We summarize below the pattern of symmetry breaking

\begin{center}
\tikzset{every picture/.style={line width=0.75pt}} 

\begin{tikzpicture}[x=0.75pt,y=0.75pt,yscale=-1,xscale=1]

\draw [color={rgb, 255:red, 202; green, 1; blue, 1 }  ,draw opacity=1 ]   (324.28,90.4) -- (324.28,114.91) ;
\draw [shift={(324.28,116.91)}, rotate = 270] [color={rgb, 255:red, 202; green, 1; blue, 1 }  ,draw opacity=1 ][line width=0.75]    (6.56,-1.97) .. controls (4.17,-0.84) and (1.99,-0.18) .. (0,0) .. controls (1.99,0.18) and (4.17,0.84) .. (6.56,1.97)   ;
\draw [color={rgb, 255:red, 248; green, 150; blue, 30 }  ,draw opacity=1 ]   (324.27,159.2) .. controls (325.99,160.81) and (326.04,162.48) .. (324.43,164.2) .. controls (322.82,165.92) and (322.87,167.58) .. (324.59,169.19) .. controls (326.31,170.8) and (326.37,172.47) .. (324.76,174.19) .. controls (323.15,175.91) and (323.2,177.58) .. (324.92,179.19) .. controls (326.64,180.8) and (326.7,182.47) .. (325.09,184.19) .. controls (323.48,185.91) and (323.53,187.57) .. (325.25,189.18) .. controls (326.97,190.79) and (327.02,192.46) .. (325.41,194.18) -- (325.54,198.01) -- (325.8,206) ;
\draw [shift={(325.87,208)}, rotate = 268.12] [color={rgb, 255:red, 248; green, 150; blue, 30 }  ,draw opacity=1 ][line width=0.75]    (7.65,-3.43) .. controls (4.86,-1.61) and (2.31,-0.47) .. (0,0) .. controls (2.31,0.47) and (4.86,1.61) .. (7.65,3.43)   ;
\draw [color={rgb, 255:red, 202; green, 1; blue, 1 }  ,draw opacity=1 ]   (200.27,84.27) -- (300.53,115.05) ;
\draw [shift={(302.44,115.64)}, rotate = 197.07] [color={rgb, 255:red, 202; green, 1; blue, 1 }  ,draw opacity=1 ][line width=0.75]    (6.56,-1.97) .. controls (4.17,-0.84) and (1.99,-0.18) .. (0,0) .. controls (1.99,0.18) and (4.17,0.84) .. (6.56,1.97)   ;

\draw (157.2,46.73) node [anchor=north west][inner sep=0.75pt]  [font=\normalsize]  {$SU( 2N_{f}) \times U( 1)_{A} \times U( 1)_{P}{}_{Q}\overset{\textcolor[rgb]{0.39,0.05,0.08}{2N}\textcolor[rgb]{0.39,0.05,0.08}{_{f}^{2}}\textcolor[rgb]{0.39,0.05,0.08}{-N}\textcolor[rgb]{0.39,0.05,0.08}{_{f}}\textcolor[rgb]{0.39,0.05,0.08}{+1}}{\rightsquigarrow } Sp( 2N_{f}) \ $};
\draw (255.4,128.93) node [anchor=north west][inner sep=0.75pt]  [font=\normalsize]  {$\textcolor[rgb]{0.79,0,0}{SU}\textcolor[rgb]{0.79,0,0}{(}\textcolor[rgb]{0.79,0,0}{N}\textcolor[rgb]{0.79,0,0}{_{f}}\textcolor[rgb]{0.79,0,0}{)}\textcolor[rgb]{0.79,0,0}{_{V}}\textcolor[rgb]{0.79,0,0}{\times U}\textcolor[rgb]{0.79,0,0}{(}\textcolor[rgb]{0.79,0,0}{1}\textcolor[rgb]{0.79,0,0}{)}\textcolor[rgb]{0.79,0,0}{_{B}} \ $};
\draw (333.28,90.75) node [anchor=north west][inner sep=0.75pt]    {$\textcolor[rgb]{0.79,0,0}{\mu }$};
\draw (288.6,216.4) node [anchor=north west][inner sep=0.75pt]  [font=\normalsize]  {$Sp( N_{f})_{V} \ $};
\draw (340.6,163.73) node [anchor=north west][inner sep=0.75pt]  [font=\footnotesize]  {$\textcolor[rgb]{0.97,0.59,0.12}{\frac{N_{f}^{2} -N_{f}}{2}}$};
\draw (192.2,163.33) node [anchor=north west][inner sep=0.75pt]  [font=\small] [align=left] {superfluid phase};
\draw (201.2,88.1) node [anchor=north west][inner sep=0.75pt]  [font=\footnotesize,rotate=-16.73]  {$\textcolor[rgb]{0.79,0,0}{m,\ a}\textcolor[rgb]{0.79,0,0}{,\ a}\textcolor[rgb]{0.79,0,0}{_{P}{}}\textcolor[rgb]{0.79,0,0}{_{Q}}$};

\end{tikzpicture}
\end{center}

From the picture is clear that the spectrum of Goldstones remains the same as in the case without the axion with an additional singlet massive state.

 \bigskip
In the next section, we will study the dispersion relations of the light modes describing the infrared dynamics by explicitly expanding the Lagrangian at the quadratic order in the fluctuations around the ground state.

\subsection{Fluctuations spectrum}
\label{FSDR}

To analytically determine the dispersion relations of the different relevant states  around the  vacuum we consider the large $a/m_\pi^2$ expansion and stop at the leading order. We recover eqs. (69) and (84) of \cite{kogut2000qcd}  for $\theta=0$, and generalize them to include the $\theta$-angle by taking it into account via an effective quark mass matrix. We obtain
\begin{align}
    \omega_1^2 &= k^2 + \mu^2\,, \qquad & {\small{ \yng(2)}} \qquad & \frac12 N_f(N_f+1)   \label{pa}\\
    \omega_2^2 &=k^2 + \frac{m_{\pi }^4 X^2}{\mu ^2 N_f^2} \,, \qquad & {\small{ \yng(1,1)}}  \qquad & \frac12 N_f(N_f-1)-1  \\
   \omega_3^2 &=k^2 + \frac{2 \left(\mu ^4 N_f^2+3 m_{\pi }^4
  X^2\right)}{ N_f^2 \mu
   ^2}+ A \,, \qquad & \bullet + {\small{ \yng(1,1)}}  \qquad & \frac12 N_f(N_f-1) \\
    \omega_4^2 &=k^2 +  \frac{2 \left(\mu ^4 N_f^2+3 m_{\pi }^4
  X^2\right)}{ N_f^2 \mu
   ^2}-A \,, \qquad & \bullet + {\small{ \yng(1,1)}}  \qquad & \frac12 N_f(N_f-1) \label{gol} \\
   \omega_5^2 &=k^2+M^2_{S}  \,, \qquad & \bullet  \qquad & 1
\end{align}
where
\begin{align}
A &= \frac{2 }{N_f^2 \mu^2}  \sqrt{\left(N_f^2 \mu
   ^4 +3 m_{\pi }^4
  X^2\right)^2+4 N_f^2
   \mu ^2 m_{\pi }^4 k^2 X^2} \,, \\ M^2_{S} &=\frac{a \mu ^4 N_f^3+2  \mu ^2 m_\pi^4 X^2}{2 \mu ^4 N_f^2- 2 m_\pi^4 X^2}\left(1-\frac{m_{\pi }^4 X^2}{\mu ^2 N_f^2} \right)\ . \label{metanon} 
\end{align}

The Young tableaux describe the irreducible representations of $Sp(N_f)$ according to which the states transform with $\bullet$ denoting the singlet representation. The number of degrees of freedom sum to $\text{dim}\left(\frac{U(2N_f)}{Sp(2N_f)} \right)=N_f(2N_f-1)$, i.e the number of pseudo-Goldstones of the chiral symmetry breaking plus the $\eta^\prime$ like $S$-particle. For $\theta=0$ the first four dispersion relations reduce to the ones found in \cite{kogut2000qcd}. $\omega_4$ describes true Goldstone modes with speed $v_G=1$ which parametrize the coset $\frac{SU(N_f)}{Sp(N_f)}$ and correspond to the $\pi$ modes considered in \cite{Orlando:2020yii}. These are the modes in which we will be mainly interested in the second part of this work since they dominate the large-charge dynamics. In addition, we have modes with a mass of order $\mu$ ($\omega_1$ and $\omega_3$) and modes with mass $\frac{m_\pi^2 X}{\mu N_f}$. 
If we include the axion in our theory, the dispersion relations of the modes described by eqs.\eqref{pa}-\eqref{gol} remain the same with $X=N_f$, i.e. we have the spectrum at $\theta=0$ as expected from the Peccei-Quinn mechanism.
On top of that, we have the modes stemming from the mixing between the singlet $S$ and the axion, which is described by the following inverse propagator
\begin{align}
 D^{-1}= \left(\begin{array}{ccc}
 \frac{\omega^{2}-k^{2}}{\sin^2\varphi} -M_{S}^{2} & -\frac{a \sqrt{N_f} a_{PQ}}{4 \sqrt{2} \nu_{PQ} \sin^2\varphi}\\
 -\frac{a \sqrt{N_f} a_{PQ}}{4 \sqrt{2} \nu_{PQ} \sin^2\varphi} & \frac{\omega^{2}-k^{2}}{4 \nu^2 \sin^2\varphi} -M_{\hat{a}}^{2}
\end{array}\right)\ ,
\end{align}
where
\begin{align}
    M^2_{S}=&\frac{\left(a \mu ^4 N_f+2 \mu^2 m_\pi^4\right)}{2\mu ^4-2 m_\pi^4}\\
    M^2_{\hat{a}}=&\frac{a \mu ^4 a_{PQ}^2}{16 \nu_{PQ}^2 \left(\mu ^4-m_\pi^4\right)}\ .
\end{align}
The dispersion relations read
\begin{align}
\label{omegaNCa1}
 \omega_{6,7}=&\frac{1}{2} \sqrt{4 k^2+8 \nu ^2 M^2_{\hat{a}} \sin^2\varphi+2 M^2_{S} \sin^2\varphi \pm \frac{1}{\nu_{PQ}}\sqrt{2 a^2 \nu ^2 N_f a_{PQ}^2+4 \nu_{PQ}^2 \sin^4\varphi\  \left(M^2_{S}-4 \nu ^2 M^2_{\hat{a}}\right)^2}} \ . 
\end{align}
As can be seen from the inverse propagator, the $S$ particle and the axion decouple in the anomaly-free limits $a \to 0$ and $a_{PQ} \to 0$. In the former case, the $S$-particle describes a pseudo-Goldstone mode with mass of order $M_S^2=\frac{\mu^2 m_\pi^4}{\mu^4-m_\pi^4}$, while for $a_{PQ} \to 0$ we have $M_S^2= \frac{a N_f}{2}$ for $\mu \gg m_\pi$, in agreement with the Witten-Veneziano relation \cite{witten1979current,veneziano1979u}. On the other hand, in the limit of vanishing anomalies, the axion is  massless being the $U(1)_{PQ}$ symmetry exact.

\section{Conclusions}
\label{Conclusions}

We investigated the spectrum, the pattern of chiral symmetry breaking and the dispersion relations at low energy of two-color QCD as a function of the baryon chemical potential, the topological term responsible for strong $CP$ breaking as well as the quark masses. The analysis is applicable to $Sp(2N)$ gauge groups with fermions in the fundamental representation. We explicitly considered the dynamics stemming from two, three, six, seven and eight light matter flavors and determined the normal and superfluid ground state. We showed that the vacuum acquires a rich structure stemming from the presence of the  $CP$ violating topological operator and unveiled novel phases. We analysed these phases, studied the dependence of the critical chemical potential on the $\theta$ angle, delineating the boundary between the normal phase and the superfluid phase. By investigating the behaviour of the $CP$  order parameter we  characterised the order of the phase transitions which are shown to be first order.  The results are readily applicable, in the normal phase, to the $\theta$ and axion physics of composite Goldstone Higgs models. In particular, a new composite strong source of $CP$ violation can be relevant to investigating composite electroweak baryogenesis \cite{Cline:2008hr,Ryttov:2008xe,Bruggisser:2018mrt}  while the nonzero chemical potential analysis is useful for studying asymmetric dark matter \cite{Ryttov:2008xe,Hochberg:2014kqa,Hansen:2015yaa} and the interplay with $CP$ violation. Our predictions can guide and be tested by first principle lattice simulations at nonzero baryon chemical potential but including also the effects of the topological susceptibility \cite{DelDebbio:2004ns,Alles:2006ur}. This work lays the foundation for our Part II in which we extend the framework to include a dilaton state to be able to approach the conformal window. In this case we will be using the large charge \cite{Hellerman:2015nra, Gaume:2020bmp, Antipin:2020abu} approach to obtain near-conformal information for the conformal dimensions of fixed baryon charge operators going beyond  the initial investigation performed in \cite{Orlando:2020yii}. 

\section*{Acknowledgments}

M. T. was supported by Agencia Nacional de Investigación y Desarrollo (ANID) grant 72210390.

\printbibliography
\end{document}